\documentclass[reprint, superscriptaddress, amsmath, amssymb, aps]{revtex4-1}
\usepackage{graphicx}
\usepackage{amsmath}
\usepackage{amssymb}
\usepackage{hyperref}
\usepackage{subcaption}
\usepackage{xcolor}
\begin{document}

\title{Double-ionization mechanisms of magnesium driven by electron impact}

\author{S. Mittal}
\affiliation{School of Physics, Georgia Institute of Technology, Atlanta, Georgia 30332-0430, USA}
\author{J. Dubois}
\affiliation{Aix Marseille Univ, CNRS, Centrale Marseille, I2M, Marseille, France}
\affiliation{Max Planck Institute for the Physics of Complex Systems, Dresden, Germany}
\author{C. Chandre}
\affiliation{Aix Marseille Univ, CNRS, Centrale Marseille, I2M, Marseille, France}
\author{T. Uzer}
\affiliation{School of Physics, Georgia Institute of Technology, Atlanta, Georgia 30332-0430, USA}

\begin{abstract}
We study double ionization of Mg by electron impact through the vantage point of classical mechanics. We consider all electron-electron correlations in a Coulomb four-body problem, where three electrons belong to the atom and the fourth electron causes the impact ionization. From our model we compute the double-ionization probability of Mg for impact energies from 15, to 125~eV. Double ionization occurs through eight double-ionization mechanisms, which we classify into four categories: inner shell capture, direct, delay and ionized inner shell mechanisms. We show that delay and ionized inner shell mechanisms require electron-electron correlations among the four electrons, and are responsible for the second increase in the double-ionization probability. Furthermore, we show that our theoretical prediction about the relative prominence of certain double ionization mechanisms is in agreement with experimental results on the relative prominence of non-first- over first-order mechanisms.
\end{abstract}

\maketitle

\section{Introduction}

Multiple ionization by an electron impact is a phenomenon wherein a free electron collides with bound electrons in a target (an atom, a molecule, an ion and so forth) and ionizes them. This phenomenon is used to study energy exchanges and electron-electron correlations in many-electron systems, electron dynamics and the structure of the target. Multiple ionization of atoms by electron is also relevant in X-ray astronomy because it is a cosmic source of X-rays~\cite{Kallman2007}. Knowledge about multiple ionization of atoms by electron impact is useful for modelling in fusion plasma~\cite{Bolt2002}. Moreover, double ionization of atoms by electron impact may help interpreting photon-atom interactions~\cite{Samson1990,Wang1995}.
\par
\begin{figure}[pt!]
    \includegraphics{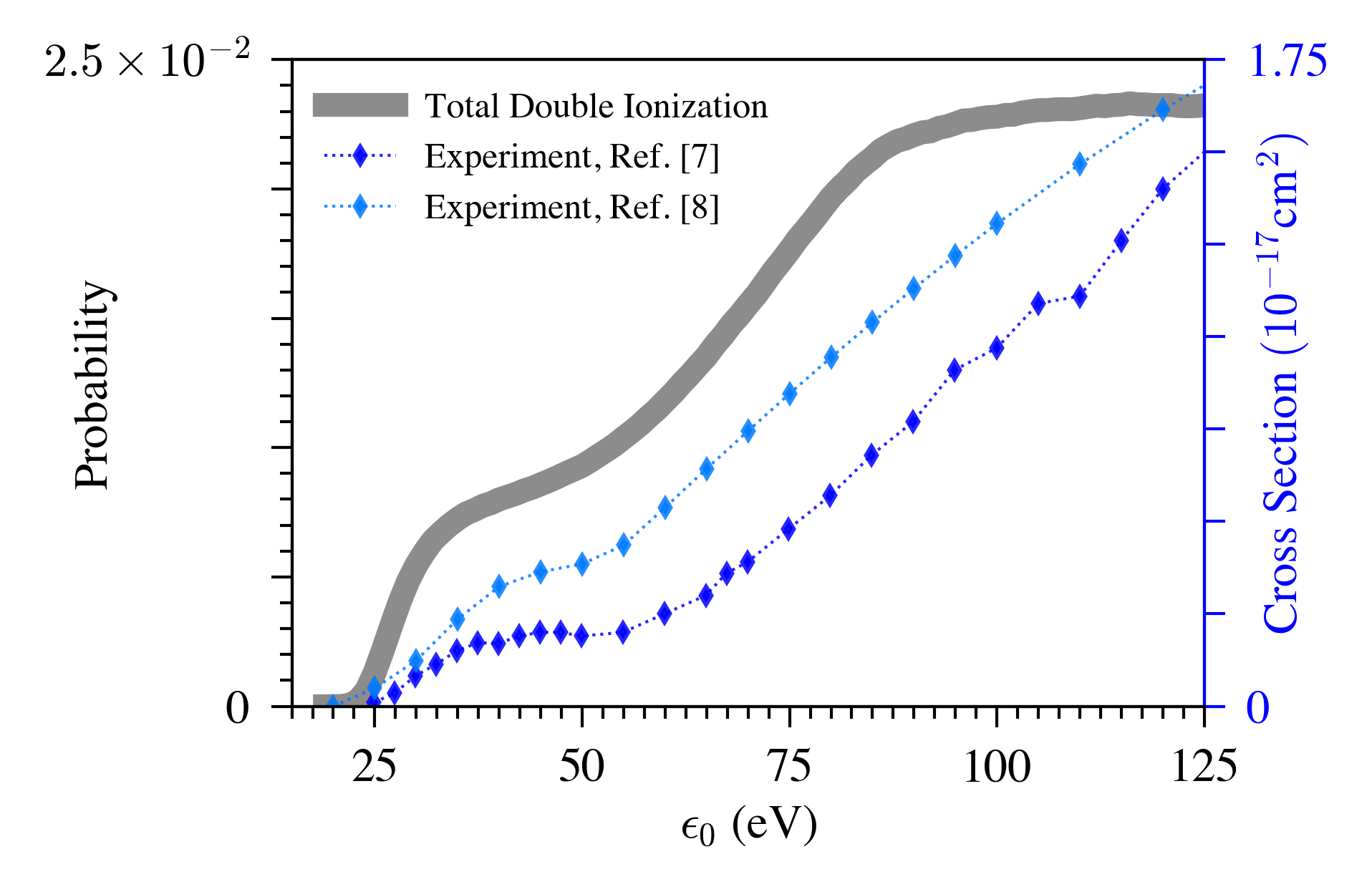}
    \caption{The solid dark gray line depicts the total probability of double ionization of Mg by electron impact found in our theoretical study. The probability is measured along the left vertical axis. The diamonds with dotted lines in shades of blue (gray) depict two experimental cross sections for double ionization of Mg by electron impact: data from Ref.~\cite{McCallion1992} is in dark blue (dark gray) and data from Ref.~\cite{Boivin1998} is in light blue (light gray). The cross-section is measured along the right vertical axis.}
\label{fig:comparisonwithexperiment}
\end{figure}

In this study, we consider the double ionization of Mg by electron impact. The probability of double ionization of Mg by electron impact was experimentally measured in Ref.~\citep{Okudaira1970,Karstensen1978,McCallion1992,Boivin1998}. In all the experiments, two drastic increases are observed in the probability of double ionization as a function of the initial energy of the impact electron (impact energy) as observed in Fig.~\ref{fig:comparisonwithexperiment}. The first increase occurs for impact energies greater than 25~eV, which corresponds to the sum of the first two ionization potentials of Mg. The second increase occurs for impact energies between 40--70~eV. The second increase lends a `knee' structure to the probability of double ionization as function of the impact energy. A similar knee structure is observed in the probability of double ionization versus impact energy for Boron ion~\cite{Shevelko2005}, Barium~\cite{Dettmann1982,Jha1994}, Calcium and Strontium~\cite{Okudaira1970, Chatterjee1984}, Argon and Xenon ions~\cite{Pindzola1984,McCallion1992}, and other target species~\cite{Freund1990}. Ref.~\cite{Jonauskas2014} attributes the knee structure to the emergence of different double ionization mechanisms. In particular, Ref.~\cite{Peach1970} conjectures that the knee structure marks the emergence of the Auger effect. The conjecture, in turn, suggests the prominent role of the inner shell electrons (electrons other than the valence shell electrons) in the double ionization of targets by electron impact for impact energies greater than a target specific energy value.
\par
In principle, the double ionization of a target by electron impact is a Coulomb $n$-body problem, with $n \geqslant 3$, and with $n \geqslant 4$ to accommodate the Auger effect. Therefore, the simplest model, which accounts for the role of the inner shell electron in the double ionization of a target by electron impact, yields a Coulomb 4--body problem. That model contains one inner shell and two valence shell electrons in the target and one impact electron. A quantum mechanical description of the simplest model may reveal the electron dynamics that cause the knee structure in the probability of double ionization as a function of the impact energy. However, a complete quantum mechanical description of a Coulomb 4--body problem is computationally too complex to be solved with present technology. Therefore, several approximate quantum mechanical approaches have been developed, of which the reduced non-perturbative methods are Time-Dependent Close Coupling (TDCC), R-Matrix with Pseudo-States (RMPS), and Convergent Close Coupling (CCC) methods~\cite{Pindzola2007,Ballance2007,Pindzola2004,Bray2002}, and the perturbative method is the Distorted Wave (DW) method~\cite{Younger1980,Bote2008}. The non-perturbative methods have successfully tackled the Coulomb 3--body problem~\cite{Bray2002}. So, the present quantum approaches reduce the Coulomb 4--body problem, of the simplest model of double ionization of a target by electron impact, to a series of Coulomb 3--body problems. For instance, Ref.~\cite{Pindzola2009} selectively applies the TDCC and DW methods on different sets of electrons and in different domains of impact energies to reproduce the knee in the probability of double ionization. Similarly, Ref.~\cite{Jha2002} and Ref.~\cite{Kumari2012} perform two separate TDCC calculations on two distinct sets of electrons and add the probability contributions from the two calculations to recover the knee structure in the probability of double ionization. The piece-wise treatments in Ref.~\cite{Pindzola2009,Jha2002,Kumari2012} serve as evidence that the double ionization of a target by electron impact is indeed not a Coulomb 3--body problem, but a more complicated, fully coupled, Coulomb $n$--body problem for $n \geqslant 4$. A comparison between theory and experiment further supports the idea that correlations among more than three electrons are required to explain the probability of double ionization of a target by electron impact~\cite{Ford1998}. Therefore, the present quantum mechanical approaches are limited in that they approximate electron-electron correlations in their model of double ionization of a target by electron impact~\cite{Pindzola2011, Pindzola2009,Jha2002,Kumari2012}.
\par
Classical mechanics provides a complementary approach to the present quantum mechanical approaches. Ref.~\cite{Gryzinski1959,Gryzinski1965,Gryzinski1999} treat electron-target collisions in the framework of classical mechanics and derive the formulas for the probability of double ionization of targets by electron impact. The results agree qualitatively and quantitatively well with experimental results. The classical mechanical approach is not only computationally more efficient than quantum mechanical approaches, but also presents two advantages over quantum mechanical approaches. First, classical mechanical approaches describe electron dynamics locally in phase space, while quantum mechanical approaches describe electron dynamics globally in phase space. 
A corollary benefit is that increasing the system size (number of electrons in the model or the spatial dimensions) is much more advantageous in classical mechanical treatments as opposed to quantum mechanical treatments. The second advantage is that classical mechanical trajectories provide intuition for making approximations, such as the binary encounter approximation~\cite{Kumari2012,Gryzinski1999,Chatterjee1984,Chatterjee1982}, in the complementary quantum mechanical approaches. Thus, classical mechanical approaches are relevant in interpreting a truly quantum phenomenon, and we employ a classical mechanical approach to describe the double ionization of Mg by electron impact.
\par
In this article, we consider three bound electrons and a static ionic core in the target Mg atom, and one impact electron. One of the three bound electrons in the atom resides closest to and most tightly bound to the ionic core in the ground state configuration of the atom. That electron is referred to as the inner shell electron. The other two electrons are further away from and more loosely bound to the ionic core as compared to the inner shell electron. They are referred to as the valence shell electrons. We consider Coulomb 4--body problem that accounts for all electron-electron correlations among the four (one inner shell, two valence shell and one impact) electrons. In our study, we do not tune the parameters in the model to achieve quantitative agreement between the theoretical double ionization probability and experimental double ionization cross section. Instead, we keep parameters in the model fixed, which are dependent on the target atom, and look only for qualitative agreement between the theoretical double ionization probability and the experimental double ionization cross section. In Fig.~\ref{fig:comparisonwithexperiment}, we depict the theoretical double ionization probability found in this study by the solid dark gray line, and we depict two experimental double ionization cross sections from Refs.~\cite{McCallion1992,Boivin1998} with diamonds with dotted line in shades of blue (gray). The correlation between the two trends in Fig.~\ref{fig:comparisonwithexperiment} confirms that our model achieves a qualitative agreement with experiments. In particular, our purely classical mechanical treatment of the problem reproduces the knee structure in the probability of double ionization of Mg by electron impact. In this article, we use classical trajectories to illustrate and understand the mechanisms behind the knee structure.
\par
In Sec.~\ref{sec:model}, we introduce the Hamiltonian model for the double ionization of Mg by electron impact and the choice of interaction potentials. In Sec.~\ref{sec:probability_curves}, we present the resulting probability of double ionization of Mg by electron impact. Next, we describe and illustrate the double ionization mechanisms, and the change in the contribution of the different mechanisms to the probability of double ionization. In Sec.~\ref{sec:discussion}, we conclude with two points of discussion; first, about the similarities and differences between our three-active electron target atom and a two-active electron target atom~\cite{Dubois2017} for Mg, and, second, about the experimental link of the mechanisms predicted by our model.
\section{Hamiltonian model \label{sec:model}}

In our model, the Mg atom is composed of four components: one static ionic core, one inner shell electron and two valence shell electrons. The static ionic core consists of a fixed nucleus and five dynamically frozen electrons, screening the charge of the nucleus. The three-active electrons interact among themselves and with the impact electron.

\subsection{Hamiltonian}

In what follows, all quantities are in atomic units (a.u.), unless explicitly stated otherwise. The electrons in the target atom are labeled by $i \in \lbrace 1,2,3 \rbrace$. The position of the $i$-th electron is denoted $\mathbf{r}_i$ and its canonically conjugate momentum is denoted $\mathbf{p}_i$. The Hamiltonian of the isolated target atom comprised of for a static ionic core and three-active electrons reads 
\begin{equation}
\label{eq:Hamiltonian_Mg}
H_{\textrm{Mg}} = \sum_{i = 1}^{3} \left[ {\frac{{\textbf{p}_i}^2}{2} + V_{\textrm{ec}}\left(\textbf{r}_i\right) + \sum_{j = i+1}^{3}{V_{\textrm{ee}}\left(\textbf{r}_i - \textbf{r}_j\right)}}\right].
\end{equation}
In the outer summation, the first term is the kinetic energy and the second term is the electron-core potential, $V_{\textrm{ec}}\left(\textbf{r}_i\right)$, for each electron. The second summation encapsulates the electron-electron potential, $V_{\textrm{ee}}\left(\textbf{r}_i - \textbf{r}_j\right)$, between the $i$-th and the $j$-th electrons. Initially, the Mg atom is in the ground state of energy $\mathcal{E}_g$, defined as the sum of the first three ionization potentials. 
\par
The impact electron is labeled by $i = 0$. The impact electron induces a perturbation in the target atom. The Hamiltonian for this perturbation reads
\begin{equation*}
H_{\textrm{I}} = \frac{{\textbf{p}_0}^2}{2} + V_{\textrm{ec}}\left(\textbf{r}_0\right) + \sum_{j = 1}^{3} {V_{\textrm{ee}}\left(\textbf{r}_0 - \textbf{r}_j\right)},
\end{equation*}
where the first term is the kinetic energy and the second term is the electron-core potential, $V_{\textrm{ec}}\left(\textbf{r}_0\right)$, of the impact electron. The summation encapsulates the electron-electron interactions between the impact electron and the three-active electrons in the target atom, $V_{\textrm{ee}}\left(\textbf{r}_0 - \textbf{r}_j\right)$. The initial energy of the impact electron is $H_{\rm I} = \epsilon_0$ and is referred to as the impact energy. The total Hamiltonian describing the electron dynamics is given by
\begin{equation*}
H = H_{\textrm{Mg}} + H_{\textrm{I}} .
\end{equation*}
\par
Here, we consider two-spatial dimensions. We note that it is still possible to carry out the calculations in three dimensions. However, for the same number of initial conditions, calculations in three spatial dimensions yield smaller double ionization probabilities as compared to calculation in two spatial dimensions. Therefore, to achieve a good resolution in the trend for probability of double ionization as a function of $\epsilon_0$, a greater number of initial conditions is required, which makes the calculations computationally more expensive. In addition to that cost, the mechanisms are not as clear in three spatial dimensions as they are in two spatial dimensions because of a higher dimensional phase space corresponding to three spatial dimensions.

\subsection{Choice of Potentials}

The electron-core potentials, $V_{\textrm{ec}}\left(\textbf{r}_i\right)$, and the electron-electron potentials, $V_{\textrm{ee}}\left(\textbf{r}_i - \textbf{r}_j\right)$, in the Hamiltonian in Eq.~\eqref{eq:Hamiltonian_Mg}, must satisfy the following three criteria to ensure a stable ground state Mg atom in our model:
\begin{enumerate}
\item {\em No self ionization.} Electrons should not ionize from the atom in the absence of the impact electron.
\item {\em Existence of an inner shell electron.} This condition requires that one of the three electrons in the atom remains the most tightly bound and the closest electron to the ionic core at all times. 
\item {\em Non-empty ground state.} The ground state is classically defined as the set of $(\{{\bf r}_i\},\{{\bf p}_i\})$ such that
\begin{equation}
\label{eq:Mg_initial_energy}
H_{\textrm{Mg}}(\{{\bf r}_i\},\{{\bf p}_i\}) = \mathcal{E}_{g}.
\end{equation}
A non-empty ground state demands that there exist at least one point in phase space $\left(\textbf{r}_1 , \textbf{p}_1 , \textbf{r}_2 , \textbf{p}_2 , \textbf{r}_3 , \textbf{p}_3 \right)$ which satisfies Eq.~\eqref{eq:Mg_initial_energy}.
\end{enumerate}
Conditions 1 and 3 are required to have a stable atom, and Condition 2 emerges from the experimentally observed ionization potentials. The first and second ionization potentials, $\mathcal{E}_1=-0.28$~a.u.\ and $\mathcal{E}_2=-0.55$~a.u., are rather close to each other, whereas the third ionization potential $\mathcal{E}_3=-2.95$~a.u., is much more negative than the first two ionization potentials. Therefore, as compared to the other two electrons, one of the electrons in the atom is more tightly bound to the ionic core.

\begin{figure}[pt!]
    \includegraphics{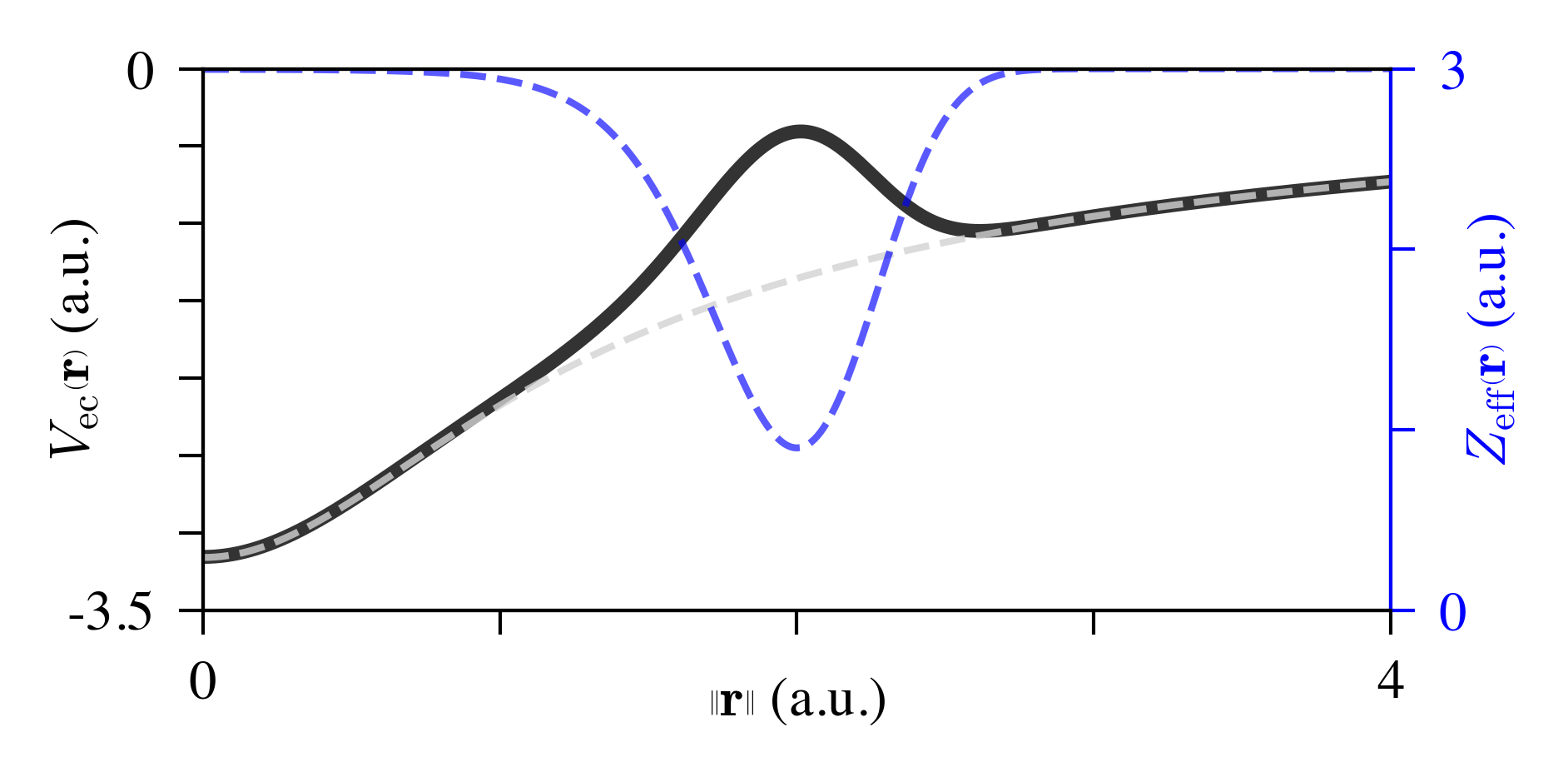}
    \caption{Effective potentials for the ion-electron interaction as a function of the distance from the ionic core. The black vertical axis on the left refers to the electron-core potential $V_{\textrm{ec}}\left(\textbf{r}\right)$, and the blue (gray) vertical axis on the right refers to the effective charge $Z_{\textrm{eff}}\left(\textbf{r}\right)$, which is given by Eq.~\eqref{eq:effective_charge} and represented by the dashed blue (dark gray) line. The ion-electron potential is given by Eq.~\eqref{eq:effective_potential}, and represented by the solid black line. The soft-Coulomb potential ${V_{\textrm{ec}}}^{\textrm{(SC)}}\left(\textbf{r}\right)$ is given by Eq.~\eqref{eq:effective_potential} and represented by the dashed light gray line.}
\label{fig:effectivepotential}
\end{figure}

\subsubsection{Soft-Coulomb potentials}

A widely used form of the electron-electron and electron-core potentials is the soft-Coulomb potentials~\cite{Becker2012,Javanainen1988}. Soft-Coulomb potentials are given by
\begin{equation}
\label{eq:SC_potential_ie}
{V_{\textrm{ec}}}^{(\textrm{SC})}(\textbf{r}) = \frac{-3}{\sqrt{\lVert\textbf{r}\rVert^2 + a^2}},
\end{equation}
and
\begin{equation}
\label{eq:SC_potential_ee}
V_{\textrm{ee}}(\textbf{r}) = \frac{1}{\sqrt{\lVert\textbf{r}\rVert^2 + b^2}}.
\end{equation}
The soft-Coulomb softening parameter $a$ controls the strength of the electron-core interaction and $b$ controls the strength of the electron-electron interaction. Soft-Coulomb potentials, given in Eq.~\eqref{eq:SC_potential_ie} and Eq.~\eqref{eq:SC_potential_ee}, might satisfy Conditions 1 and 3 for specific choices of $a$ and $b$~\cite{Ho2006}. However, the soft-Coulomb potentials do not satisfy Condition 2 for any choice of $a$ or $b$. In other words, it is possible to find non-empty ground states that do not lead to self-ionization; however, in these non-empty ground states, the inner shell electron and valence shell electrons play equal role, and are equally bound to the ionic core, which violates Condition 2. 

\subsubsection{Effective Charge \label{sec:effective_charge}}

In order to remedy this problem, we keep the electron-electron soft-Coulomb potential given in Eq.~\eqref{eq:SC_potential_ee}, and introduce an effective charge, $Z_{\textrm{eff}}\left(\textbf{r}\right)$, in place of the constant charge in the electron-core soft-Coulomb potential in Eq.~\eqref{eq:SC_potential_ie}. The resulting electron-core soft-Coulomb potential with effective charge is
\begin{equation}
\label{eq:effective_potential}
V_{\textrm{ec}}(\textbf{r}) = \frac{-Z_{\textrm{eff}}\left(\textbf{r}\right)}{\sqrt{\lVert\textbf{r}\rVert^2 + a^2}}.
\end{equation}
The role of the effective charge in the electron-core soft-Coulomb potential is to create an energy barrier, the height of which is greater than the sum of the first two ionization potentials of the atom. This energy barrier ensures that bound electrons that are on one side of the barrier can not move to the other side of the barrier by any redistribution of energy among the bound electrons in the ground state. Since the proximity of an electron to the ionic core determines the strength of the bond between the two, it follows that Condition 2 can be satisfied. We model the effective charge as
\begin{equation}
\label{eq:effective_charge}
Z_{\textrm{eff}}(\textbf{r}) = 3 + 3\left(\chi - 1\right)\exp{\left(-\alpha\left(\lVert\textbf{r}\rVert^2 - R^2\right)^2\right)},
\end{equation}
where the parameters $\chi$, $\alpha$ and $R$ govern, respectively, the height, width and position of the energy barrier. Although we choose $Z_{\textrm{eff}}(\textbf{r})$ given by Eq.~\eqref{eq:effective_charge}, we note that any function with the general features highlighted above may have been used and would lead to the same results, at least qualitatively. Figure~\ref{fig:effectivepotential} depicts the effective charge as a function of the distance from the ionic core by a dashed blue (dark gray) line. The corresponding electron-core soft-Coulomb potential, given by Eq.~\eqref{eq:effective_potential}, is depicted in solid black line. The local maximum at $\lVert\textbf{r}\rVert = 2$~a.u.\ is the peak of the energy barrier. To highlight the role of the effective charge and the energy barrier, we have depicted the soft-Coulomb potential without effective charge given by Eq.~\eqref{eq:SC_potential_ie} in dashed light gray line in Fig.~\ref{fig:effectivepotential}.
\par
In sum, the total Hamiltonian is given by
\begin{multline}
\label{eq:total_Hamiltonian}
H = \sum_{i = 0}^{3} \Bigg[\frac{{\textbf{p}_i}^2}{2} - \frac{Z_{\textrm{eff}}(\textbf{r}_i)}{\sqrt{\lVert\textbf{r}_i\rVert^2 + a^2}} \\
+ \sum_{j = i + 1}^{3}{ \frac{1}{\sqrt{\lVert\textbf{r}_i - \textbf{r}_j\rVert^2 + b^2 }}}\Bigg],
\end{multline}
with $Z_{\textrm{eff}}$ given by Eq.~\eqref{eq:effective_charge}. We choose soft-Coulomb parameters $a = 0.95$ and $b = 0.1$. Our choice allows us to find non-empty ground states and allows significant energy exchanges among the electrons. The parameters, $\chi$, $\alpha$ and $R$ are taken as 0.3, 0.4, and 2 respectively. 

\subsection{Initial Conditions}

We integrate Hamilton's equations obtained from Hamiltonian~\eqref{eq:total_Hamiltonian}, to compute trajectories. In this section, we discuss our procedure to assign initial conditions to the electrons for the integration.

\subsubsection{Impact Electron}

We start the integration at an initial time $t_{\textrm{in}}<0$.
The initial position of the impact electron is denoted by $\mathbf{r}_{0,\textrm{in}} = \hat{\mathbf{x}} x_{0,\textrm{in}} + \hat{\mathbf{y}} y_{0,\textrm{in}}$. The initial $x$-coordinate of the impact electron $x_{0,\textrm{in}}<0$ is fixed for all impact energies. The initial $y$-coordinate $y_{0, \textrm{in}}$, which is referred to as the impact parameter, is randomly chosen from a uniform distribution in $[0, 5)$~a.u.\ Initially, the impact electron is sent along the $\hat{\mathbf{x}}$-direction with impact energy $\epsilon_0$. The initial momentum of the impact electron is given by
\begin{equation*}
\mathbf{p}_{0, \textrm{in}} = \hat{\mathbf{x}} \sqrt{2\epsilon_0} .
\end{equation*}
The initial time $t_{\textrm{in}}$ is chosen such that the impact electron would reach $x=0$ at $t=0$ if the target atom were absent. Therefore,
\begin{equation*}
t_{\textrm{in}} = \frac{x_{0, \textrm{in}}}{\lVert\textbf{p}_{0, \textrm{in}}\rVert} ,
\end{equation*}
and the first collision between the impact electron and any bound electron occurs approximately at time $t = 0$.
Both $x_{0, \textrm{in}}$ and the distance beyond which an electron is considered ionized is fixed at 200~a.u.

\subsubsection{Isolated atom}

In this section, we describe the procedure to assign the initial positions and momenta to the bound electrons in the isolated atom.
\par
First, the configuration of the inner shell electron is determined in absence of valence shell electrons. The inner shell electron is randomly positioned such that $\lVert\textbf{r}_{3,\textrm{in}}\rVert < 2 \textrm{ a.u.}$ and
\begin{equation*}
V_{\textrm{ec}}\left(\textbf{r}_{3,\textrm{in}}\right) < \mathcal{E}_3,
\end{equation*}
where $\mathcal{E}_3$ is the third ionization potential of Mg. Then, the kinetic energy of the inner shell electron, $T_3$, is given by
\begin{equation*}
T_3 = \mathcal{E}_3 - V_{\textrm{ec}}\left(\textbf{r}_{3,\textrm{in}}\right).
\end{equation*}
Following which, the initial momentum of the inner shell electron becomes
\begin{equation*}
\lVert\textbf{p}_{3,\textrm{in}}\rVert = \sqrt{2T_3} .
\end{equation*}
The momenta along $\hat{\mathbf{x}}$- and $\hat{\mathbf{y}}$-direction are randomly decomposed to determine the initial momentum of the inner shell electron. 
\par
Next, we determine the configuration of the two valence shell electrons in the presence of the ionic core and the inner shell electron. The initial position of the two valence shell electrons are randomly generated such that $\lVert\textbf{r}_{1,\textrm{in}} \rVert \in [2,8]$~a.u., $\lVert\textbf{r}_{2,\textrm{in}} \rVert \in [2,8]$~a.u.\ and
\begin{multline*}
V_{\textrm{ec}}\left(\textbf{r}_{1,\textrm{in}}\right) + V_{\textrm{ec}}\left(\textbf{r}_{2,\textrm{in}}\right) + V_{\textrm{ee}}\left(\textbf{r}_{1,\textrm{in}} - \textbf{r}_{2,\textrm{in}}\right) \\ + V_{\textrm{ee}}\left(\textbf{r}_{1,\textrm{in}} - \textbf{r}_{3,\textrm{in}}\right) + V_{\textrm{ee}}\left(\textbf{r}_{2,\textrm{in}} - \textbf{r}_{3,\textrm{in}}\right) < \mathcal{E}_1 + \mathcal{E}_2,
\end{multline*}
where $\mathcal{E}_1$ and $\mathcal{E}_2$ are the first two ionization potentials, respectively. Then, the sum of the kinetic energies of the two valence shell electrons, $T_1 + T_2$, becomes
\begin{multline*}
T_1 + T_2 = \mathcal{E}_1 + \mathcal{E}_2 - \Big[ V_{\textrm{ec}}\left(\textbf{r}_{1,\textrm{in}}\right) + V_{\textrm{ec}}\left(\textbf{r}_{2,\textrm{in}}\right) \\ + V_{\textrm{ee}}\left(\textbf{r}_{1,\textrm{in}} - \textbf{r}_{2,\textrm{in}}\right) + V_{\textrm{ee}}\left(\textbf{r}_{1,\textrm{in}} - \textbf{r}_{3,\textrm{in}}\right) + V_{\textrm{ee}}\left(\textbf{r}_{2,\textrm{in}} - \textbf{r}_{3,\textrm{in}}\right)\Big] .
\end{multline*}
The initial momenta of the two valence shell electrons are determined using a parametrization of the 4-sphere by three randomly chosen angles such that
\begin{equation*}
    \lVert\textbf{p}_{1, \textrm{in}}\rVert^2 + \lVert\textbf{p}_{2, \textrm{in}}\rVert^2= 2\left(T_1 + T_2\right).
\end{equation*}

\subsection{Integration scheme}

We employ the $4^{\textrm{th}}$ order Runge-Kutta numerical integration scheme. We fix the ending time of integration at $t_{\textrm{f}} = 1500$~a.u.\ for all impact energies. We integrate $3\times10^{7}$ initial conditions for each $\epsilon_0 \in \left\{10, 11, 12, ..., 125\right\}$~eV. We fix the step size in the integration scheme at 0.1~a.u.\ Further reducing the time step size presents no perceptible changes in the double ionization probability and increases the computational time. The integration scheme was developed in C/C++ and CUDA C, and deployed on NVIDIA Tesla V100 GPU and Intel Core i9 8950HK CPU. The double ionization probability was calculated from the results of the computation done on the GPU.

\section{Double ionization probability \label{sec:probability_curves}}

\begin{figure}[pt!]
    \begin{subfigure}[]{\columnwidth}
        \includegraphics[trim = {0 1.55in 0 0} ,clip]{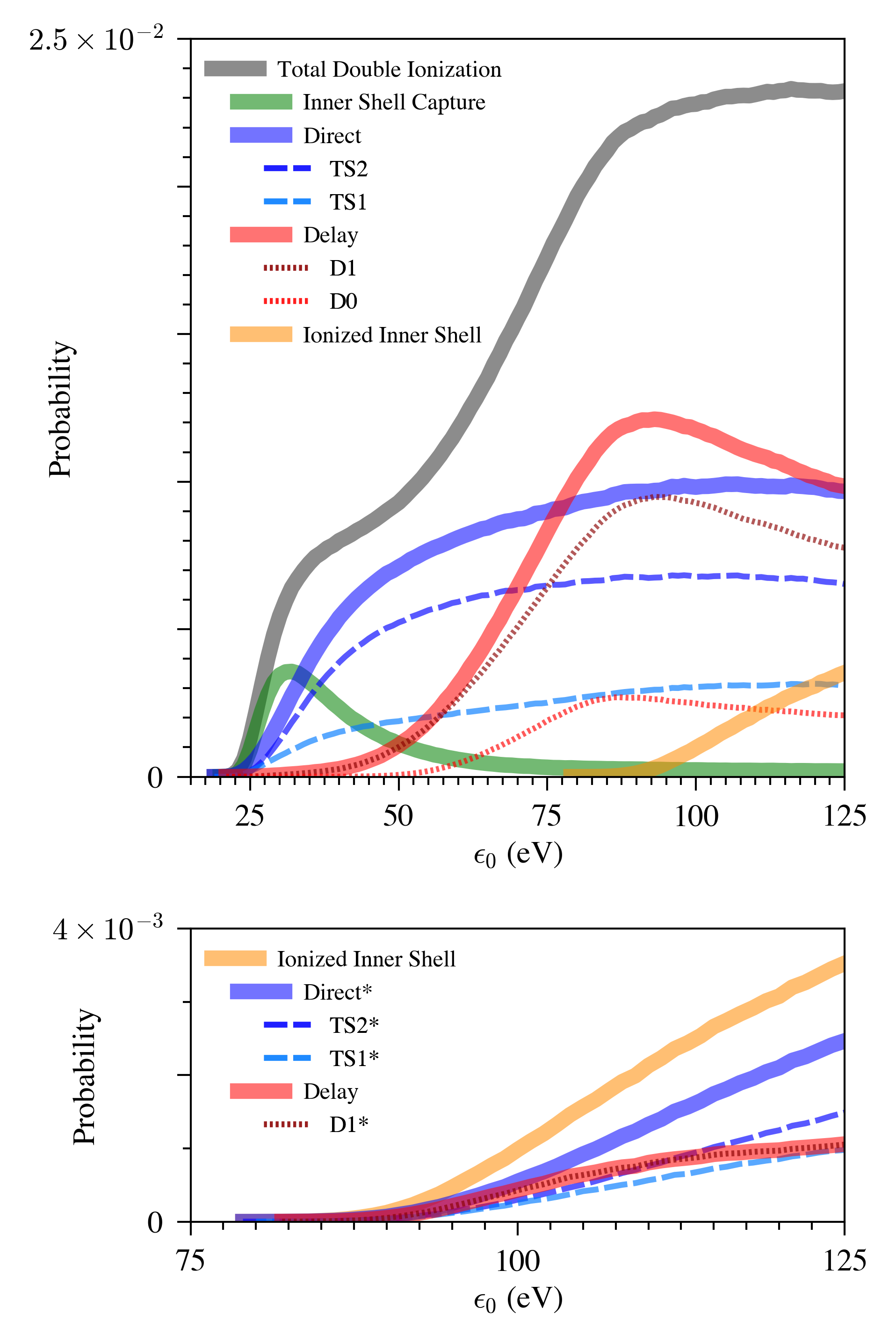}
        \caption{\label{fig:detailedcrosssectionupperpanel}\textit{Upper panel}: The double ionization probability is represented by a thick dark gray line (first thick line from the top), and the probabilities of the four main categories of mechanisms are represented by thick color lines: \textit{Inner Shell Capture} in green (thick line with peak around 30~eV), \textit{Direct} in blue (second thick line from the top around 45~eV), \textit{Delay} in red (second thick line from the top around 90~eV) and \textit{Ionized Inner Shell} in yellow (second thick line from bottom around 125~eV). The probabilities of the main mechanisms within the four main categories are depicted as follows: \textit{TS2} by dashed dark blue line (dashed dark gray line) and \textit{TS1} by dashed light blue line (dashed light gray line) under the \textit{Direct} mechanisms, and \textit{D1} by dotted dark red line (dotted dark gray line) and \textit{D0} by dotted light red line (dotted light gray line) under the \textit{Delay} mechanisms.}
    \end{subfigure}
    \begin{subfigure}[]{\columnwidth}
        \includegraphics[trim = {0 0 0 3.18in} ,clip]{figure3.png}
        \caption{\label{fig:detailedcrosssectionlowerpanel}\textit{Lower panel}: The probability of the \textit{Ionized Inner Shell} mechanisms is represented by thick yellow line (first thick line from the top). The probabilities of the associated two mechanisms are represented by thick color lines: \textit{Direct*} in blue (second thick line from the top) and \textit{Delay*} in red (first thick line from the bottom). The probabilities of the mechanisms within the \textit{Direct*} and \textit{Delay*} categories are illustrated as follows: \textit{TS2*} by dashed dark blue line (dashed dark gray line) and \textit{TS1*} in dashed light blue line (dashed light grey line) under \textit{Direct*} mechanism, and \textit{D1*} by dotted dark red line (dotted dark gray line) for the \textit{Delay*} mechanisms.}
    \end{subfigure}
    \caption{Double ionization probability for Hamiltonian~\eqref{eq:total_Hamiltonian}.}
    \label{fig:detailedcrosssection}
\end{figure}

The probability of double ionization of Mg by electron impact as a function of the impact energy $\epsilon_0$ computed from our model is represented by the thick dark gray line (first thick line from the top) in Fig.~\ref{fig:detailedcrosssectionupperpanel}. We observe that our model reproduces the knee structure in the double ionization probability for low impact energies as observed in experiments~\cite{Okudaira1970,Karstensen1978,McCallion1992,Boivin1998}.
\par
Classical mechanics allows visualizing trajectories, which unveil the different double ionization mechanisms. The probability of different mechanisms sums up to the total double ionization probability. Our model predicts that the double ionization of Mg by electron impact proceeds through eight main mechanisms in the studied range of impact energies. We classify those eight mechanisms into four categories: \textit{Inner Shell Capture}, \textit{Direct}, \textit{Delay} and \textit{Ionized Inner Shell}. We show the probabilities of the four categories of mechanisms by thick color (gray) lines in Fig.~\ref{fig:detailedcrosssectionupperpanel}: \textit{Inner Shell Capture} in green (thick line with peak around 30~eV), \textit{Direct} in blue (second thick line from the top around 45~eV), \textit{Delay} in red (second thick line from the top around 90~eV) and \textit{Ionized Inner Shell} in yellow (second thick line from bottom around 125~eV). In the \textit{Inner Shell Capture} mechanisms, the impact electron gets momentarily trapped in the inner shell region and ionizes later. As expected, these mechanisms occurs for low impact energies. The \textit{Direct} mechanisms are characterized by the absence of a time delay between impact and ionization. Roughly speaking, the \textit{Direct} mechanisms proceed via collisions like the collisions of balls in the game of billiards. On the contrary, the \textit{Delay} mechanisms are characterized by a time delay between impact (around $t=0$) and ionization. Consequently, an intermediate metastable state of the target forms during the period of delay. The last but not the least, the \textit{Ionized Inner Shell} mechanisms group together all the mechanisms in which the inner shell electron is ionized. 
\par
At about $\epsilon_0 = 50$~eV, we observe a drastic increase of the probability of \textit{Delay} mechanisms, depicted with the thick red line (second thick line from the top around 90~eV). This increase is responsible for the increase in the probability of double ionization. Then, at about $\epsilon_0 = 100$~eV, the increase in the probability of the \textit{Ionized Inner Shell} mechanisms, represented by the thick yellow line (second thick line from bottom around 125~eV), counters the decrease in the probability of the \textit{Delay} mechanisms, creating a plateau in the probability of double ionization till $\epsilon_0 = 125$~eV. 
\par
In summary, the knee shape in the double ionization probability is mostly due to the cross-over between the \textit{Inner Shell Capture} and the \textit{Delay} and \textit{Direct} mechanisms. Even if there are no visible qualitative changes in the double ionization probability for impact energies around 100~eV and higher, the probability curve is hiding the appearance of additional mechanisms such as the \textit{Ionized Inner Shell} mechanisms.
\par

\begin{figure}[pt!]
    \includegraphics{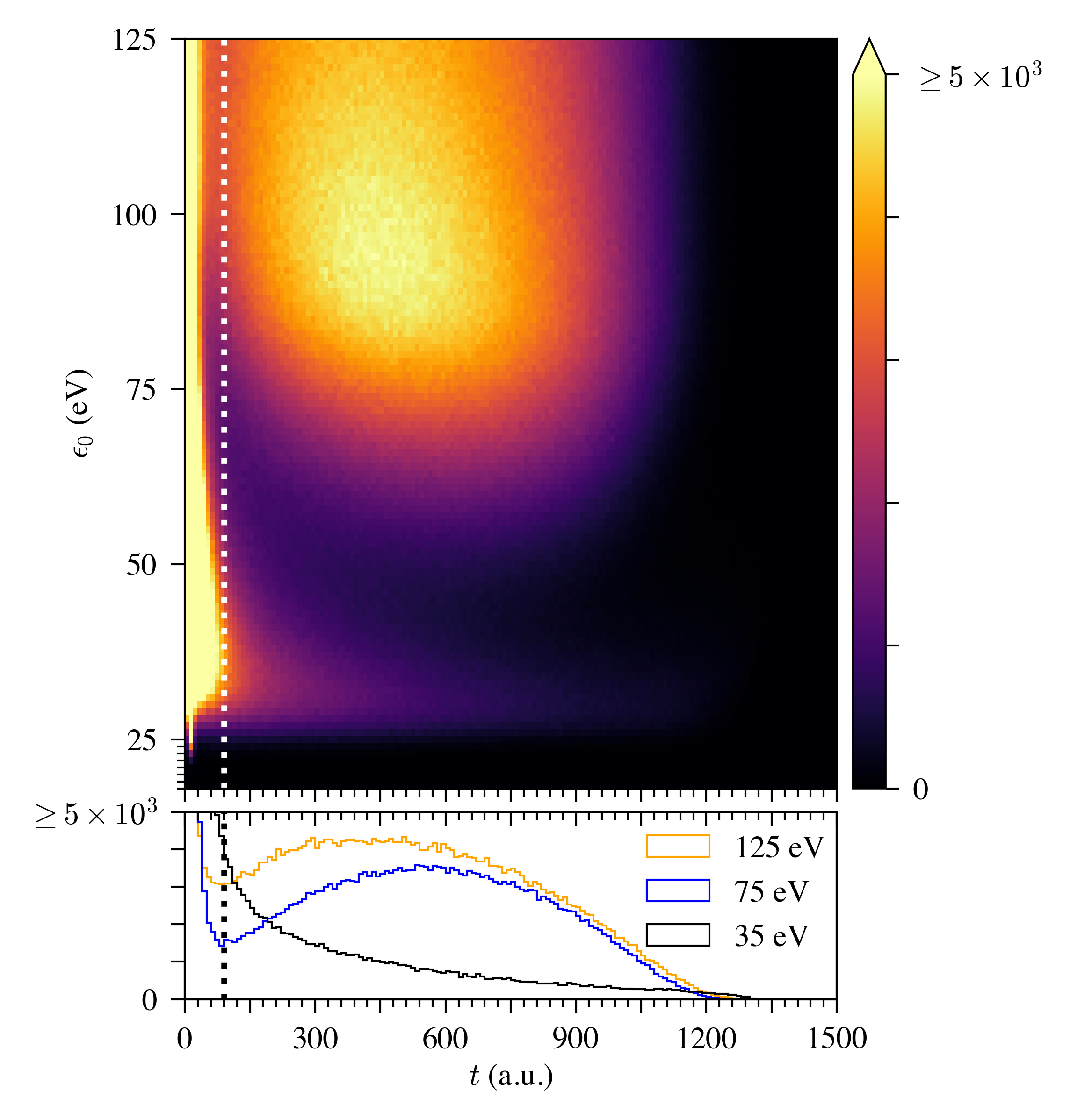}
    \caption{\textit{Upper panel}: The distribution of times when the last ionization occurs in the double ionization process as a function of the impact energy, $\epsilon_0$. We say that the last ionization occurs when the last ionized electron crosses $|\mathbf{r}| = 8$~a.u.\ The time intervals along the  horizontal axis are 10~a.u.\ wide. The color-bar indicates the count of total double ionizations for the corresponding $\epsilon_{0}$ and ionization time. \textit{Lower panel}: The distribution in times of the initiation of the last ionization for $\epsilon_{0}$ equal to 35~eV in black, 75~eV in blue (dark gray) and 125~eV in orange (light gray). The vertical axis measures the count of double ionizations. The time intervals along the horizontal axis are 10~a.u.\ wide. The dotted vertical line indicates $t_{\textrm{Th}} = 90$~a.u.\ }
\label{fig:timedistribution}
\end{figure}

Furthermore, we infer the times taken to complete the double ionizations from the trajectories. We plot the distribution of these times in Fig.~\ref{fig:timedistribution}. The upper panel of Fig.~\ref{fig:timedistribution} depicts that distribution as a function of time, $t$, and impact energy, $\epsilon_{0}$. The lower panel of Fig.~\ref{fig:timedistribution} depicts that distribution for 35~eV, 75~eV and 125~eV impact energies in black, blue (dark gray) and orange (light gray) lines, respectively. This distribution allows us to differentiate the three categories of double ionization mechanisms mentioned above: \textit{Inner Shell Capture}, \textit{Direct}, and \textit{Delay} mechanisms. In the upper panel, there are two distinct bright yellow (white) regions that imply a high number of occurrences of double ionizations. Correspondingly, there are two distinct peaks in the distribution in the lower panel. The peak in the distributions on the left for all impact energies corresponds to the \textit{Direct} mechanisms, and the peak on the right to the \textit{Delay} mechanisms. The residual occurrences to the right of the peak between 25 and 50~eV correspond to the \textit{Inner Shell Capture} mechanisms. We choose a threshold time, $t_{\textrm{Th}} = 90$~a.u., to best separate the different classes of mechanisms. Our choice of $t_{\textrm{Th}}$ is represented by the dotted white line and dotted black line in upper and lower panels of Fig.~\ref{fig:timedistribution}, respectively. All ionizations, which are not \textit{Inner Shell Capture} mechanisms, that occur before $t_{\textrm{Th}}$ are classified as \textit{Direct} mechanisms and that occur after $t_{\textrm{Th}}$ are classified as \textit{Delay} mechanisms. Therefore, $t_{\textrm{Th}}$ defines a characteristic time scale for delay, which is used to distinguish \textit{Direct} and \textit{Delay} mechanisms. \textit{Inner Shell Capture} mechanisms are distinguished from \textit{Direct} and \textit{Delay} mechanisms by measuring the time spent by two electrons in the inner shell region. While \textit{Ionized Inner Shell} mechanisms are segregated from other categories of mechanisms, using that \textit{Ionized Inner Shell} mechanisms involve ionization of the inner shell electron.
\par
Figure~\ref{fig:detailedcrosssection} illustrates the different contributions of the  different mechanisms within each category: \textit{TS2} and \textit{TS1} are the mechanisms within the \textit{Direct} category, and are represented by dashed dark blue (dashed dark gray) and dashed light blue (dashed light gray) lines respectively in Fig.~\ref{fig:detailedcrosssectionupperpanel}. Similarly, \textit{D1} and \textit{D0} are the sub-mechanisms within the \textit{Delay} category, and are represented by dotted dark red (dotted dark gray) and dotted light red (dotted light gray) lines, respectively in Fig.~\ref{fig:detailedcrosssectionupperpanel}. The mechanisms \textit{TS2*}, \textit{TS1*} and \textit{D1*} are associated with the \textit{Ionized Inner Shell} category, and are represented in Fig.~\ref{fig:detailedcrosssectionlowerpanel} by dashed dark blue (dashed dark gray), dashed light blue (dashed light gray) and dotted dark red (dotted dark gray) lines respectively. Each mechanism is explained in more detail below.

\subsection{Format for the figures of the mechanisms \label{sec:format_mechanisms}}

Before explaining each double ionization mechanism identified in our model, we introduce the format of Figs.~\ref{fig:innershellcapture}-\ref{fig:d1star}. In each of these figures, e.g., Fig.~\ref{fig:innershellcapture}, there is a schematic diagram in the top left corner, two trajectory plots (e.g., in Figs.~\ref{innershellcaptureleftcircle} and~\ref{innershellcapturerightcircle}), and a plot of the electron-electron potential versus time (e.g., Fig.~\ref{innershellcapturerectangle}).

\paragraph{Schematic diagram}

In the schematic diagrams, the labels `X', `V', `O' and the gray circular region correspond to the impact electron, valence shell electron, inner shell electron and the inner shell region, respectively. An arrow corresponds to an ionization process that occurs on short timescales (time shorter than a certain threshold $t_{\textrm{Th}}$), and a wiggly line corresponds to an ionization process that occurs on long timescales (time larger than $t_{\textrm{Th}}$). We choose a threshold $t_{\textrm{Th}} = 90$~a.u.\ for which there is a clear separation between the {\em Delay} and {\em Direct} processes. 

\paragraph{Panels (a) and (b)}

In panels (a), we plot the electron trajectories covering 0 to 80~a.u.\ from the ionic core, and in panels (b), the same electron trajectories are plotted in a range from 0 to 8~a.u.\ around the ionic core. In both panels, the black and yellow (light gray) lines represent the impact electron and inner shell electron trajectories, respectively. The red (medium gray) line represents the trajectory for that valence shell electron that ionizes and leaves the 8~a.u.\ boundary before the other valence shell electron. The blue (dark gray) line represents the trajectory for the other valence shell electron. We refer to an electron by its corresponding color in these figures, e.g. blue (dark gray) valence shell electron. We indicate the direction of approach of the impact electron with a black arrowhead. We plot the arrowheads to indicate the direction of the valence shell electrons in panels (b). The arrowheads in panels (b) correspond to the same instant of time for all electrons. 

\paragraph{Panel (c)}

In panels (c), we depict the six electron-electron potentials, $V_{\textrm{ee}}\left(\textbf{r}_i - \textbf{r}_j\right)$, for $i, j \in \left\{0, 1, 2, 3\right\}$, with an inset focusing on a time interval around the collision, i.e., around $t = 0$~a.u.\ The electron-electron potentials are colored as follows: $V_{\textrm{ee}}\left(\textbf{r}_0 - \textbf{r}_1\right)$ is red, $V_{\textrm{ee}}\left(\textbf{r}_0 - \textbf{r}_2\right)$ is blue, $V_{\textrm{ee}}\left(\textbf{r}_0 - \textbf{r}_3\right)$ is yellow, $V_{\textrm{ee}}\left(\textbf{r}_1 - \textbf{r}_2\right)$ is purple, $V_{\textrm{ee}}\left(\textbf{r}_1 - \textbf{r}_3\right)$ is brown, and $V_{\textrm{ee}}\left(\textbf{r}_2 - \textbf{r}_3\right)$ is green.

Lines in panels (a), (b) and (c) are thin and translucent until the time the impact electron first enters the 8~a.u.\ boundary, thick and solid until the last electron in double ionization leaves the 8~a.u.\ boundary for the last time, and thin and translucent for the time after.

Using this typology, we examine all the mechanisms leading to double ionization we found by a close inspection of the trajectories. 

\subsection{Inner Shell Capture Mechanism\label{sec:innershellcapturemechanism}}

\begin{figure}[pt!]
\centering
    \begin{subfigure}{0.49\columnwidth}
        \includegraphics[trim = {0 0 1.64in 0} ,clip]{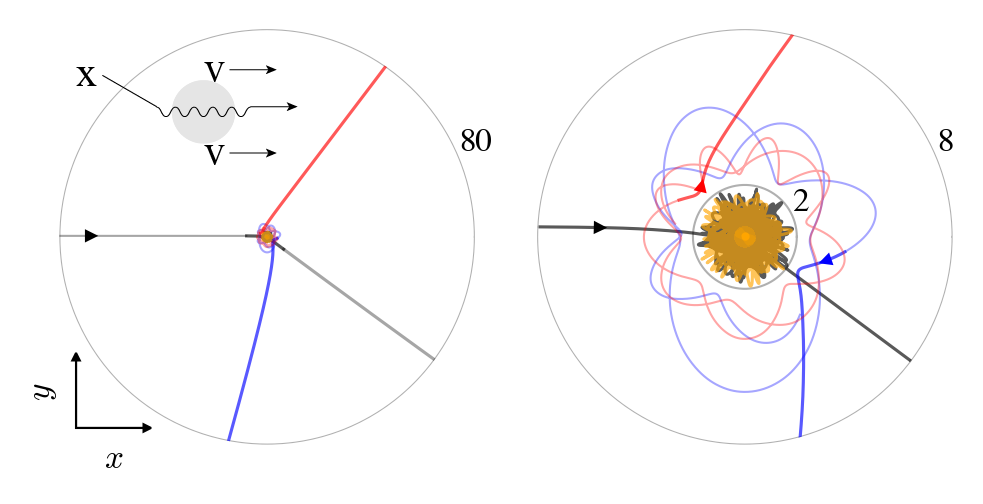}
        \caption{\label{innershellcaptureleftcircle}}
    \end{subfigure}
    \begin{subfigure}{0.49\columnwidth}
        \includegraphics[trim = {1.65in 0 0 0} ,clip]{figure5ab.png}
        \caption{\label{innershellcapturerightcircle}}
    \end{subfigure}
    \begin{subfigure}{\columnwidth}
        \includegraphics{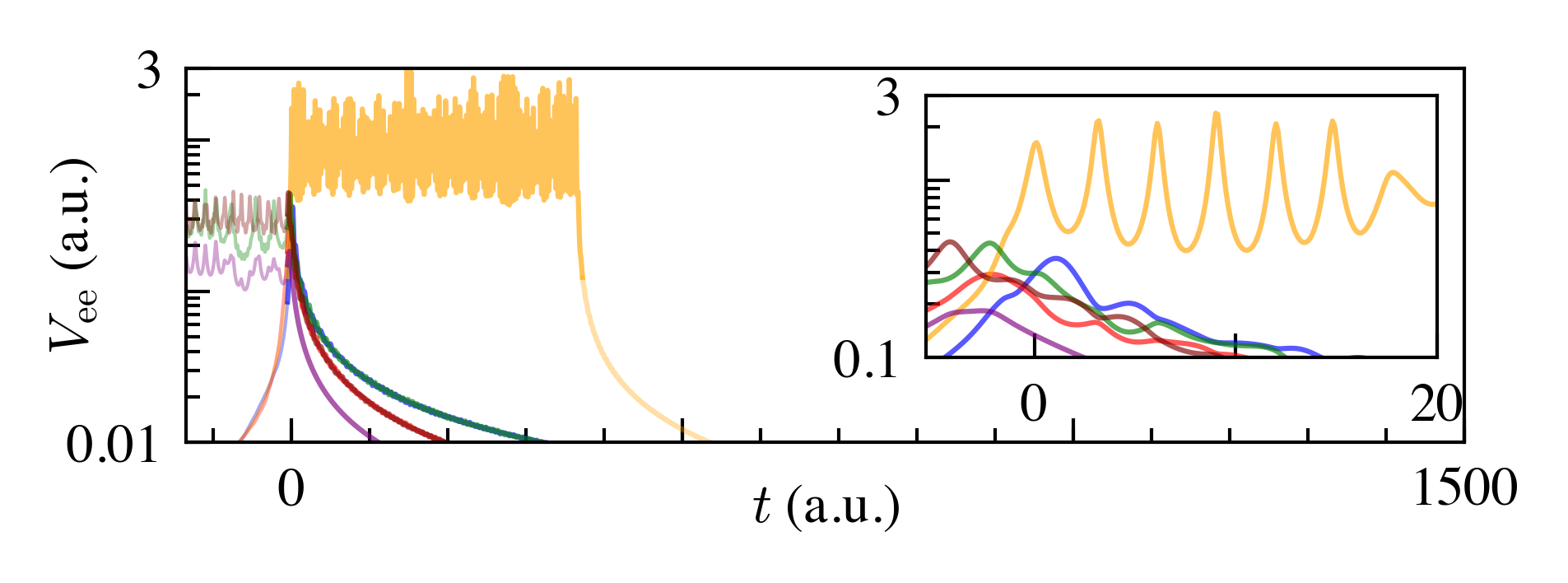}
        \caption{\label{innershellcapturerectangle}}
    \end{subfigure}
    \caption{\textit{Inner Shell Capture} mechanism at $\epsilon_{0} = 30$~eV (see Sec.~\ref{sec:format_mechanisms} for details).} 
    \label{fig:innershellcapture}
\end{figure}

An example of the \textit{Inner Shell Capture} mechanism is plotted in Fig.~\ref{fig:innershellcapture}. In Fig.~\ref{innershellcapturerightcircle}, the trajectory of the impact electron (black line) shows that the impact electron falls into the inner shell region, $\lVert\textbf{r}\rVert < 2$~a.u.\ In Fig.~\ref{innershellcapturerectangle}, the yellow (first from the top after $t = 0$) electron-electron potential shows that the impact electron and the inner shell electron interact for a time duration greater than $t_{\textrm{Th}}$. Meanwhile, the steep decrease in all but the yellow (first from the top after $t = 0$) electron-electron potential, at about time $t = 0$~a.u., implies that the two valence shell electrons leave the atom right after the impact. Finally, Fig.~\ref{innershellcaptureleftcircle} illustrates that the impact electron eventually leaves the atom, completing the double ionization mechanism. 
\par
The schematic diagram in the top left in Fig.~\ref{innershellcaptureleftcircle} summarizes this process: The impact electron `X' falls into the inner shell region (gray circle) and stays there for a long time before it leaves the atom while the two valence shell electrons `V's leave the atom before $t = t_{\textrm{Th}}$ (arrows only).
\par
The impact electron exchanges energy with the valence shell electrons, which leave the atom. During this process, the energy of the impact electron falls below the energy barrier (see Sec.~\ref{sec:effective_charge}), and the impact electron gets captured in the inner shell region. The capture persists and strengthens the electron-electron correlation between the impact electron and the inner shell electron. This mechanism is characterized by a series of energy exchanges between the impact electron and the inner shell electron, after which the impact electron finally gets enough energy to overcome the potential barrier and leaves the atom.  The \textit{Inner Shell Capture} mechanism requires an inner shell electron dynamics and the inner shell region to occur, and the mechanism proceeds through strong electron-electron correlation between the inner shell and the impact electrons.
\par
This mechanism dominates the probability of double ionization for impact energies between 10~eV and 40~eV (see Fig.~\ref{fig:detailedcrosssection}). 

\subsection{Direct Mechanism}

\textit{Direct} mechanisms involve the ionization of the two valence shell electrons before time $t = t_{\textrm{Th}}$. There are two types of \textit{Direct} mechanisms: \textit{TS1} and \textit{TS2}, which are depicted in Fig.~\ref{fig:ts1} and Fig.~\ref{fig:ts2}, respectively, and are described in this section. In the nomenclature of the mechanisms, `TS' indicates that the mechanisms lead to double ionization in two steps (TS), and `1' or `2' refers to the number of collisions between the impact electron and valence shell electrons, which trigger ionization. In \textit{TS1} mechanisms, the impact electron collides and ionizes one valence shell electron, which on its way out of the atom collides with and ionizes the other valence shell electron. While, in \textit{TS2} mechanism the impact electron collides sequentially with and ionizes both the valence shell electrons.

\subsubsection{TS1}

\begin{figure}[pt!]
\centering
    \begin{subfigure}{0.49\columnwidth}
        \includegraphics[trim = {0 0 1.64in 0} ,clip]{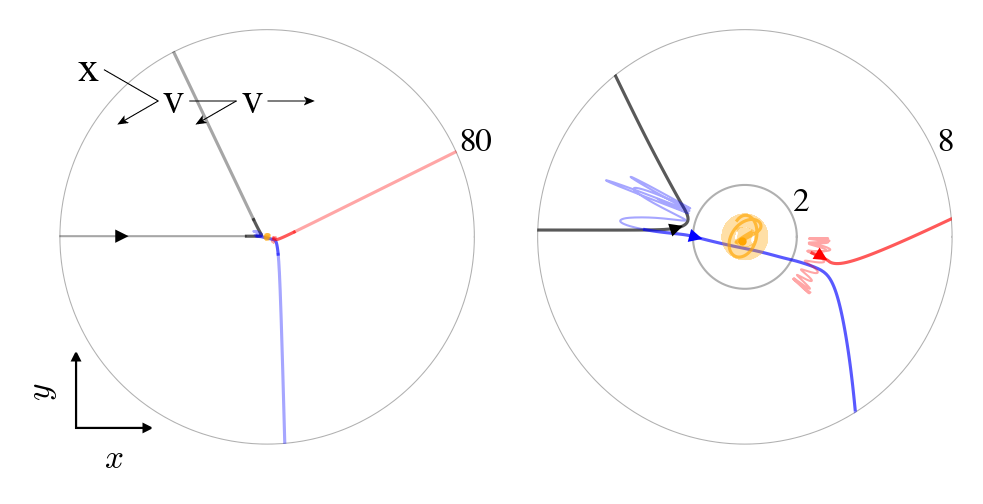}
        \caption{\label{ts1leftcircle}}
    \end{subfigure}
    \begin{subfigure}{0.49\columnwidth}
        \includegraphics[trim = {1.65in 0 0 0} ,clip]{figure6ab.png}
        \caption{\label{ts1rightcircle}}
    \end{subfigure}
    \begin{subfigure}{\columnwidth}
        \includegraphics{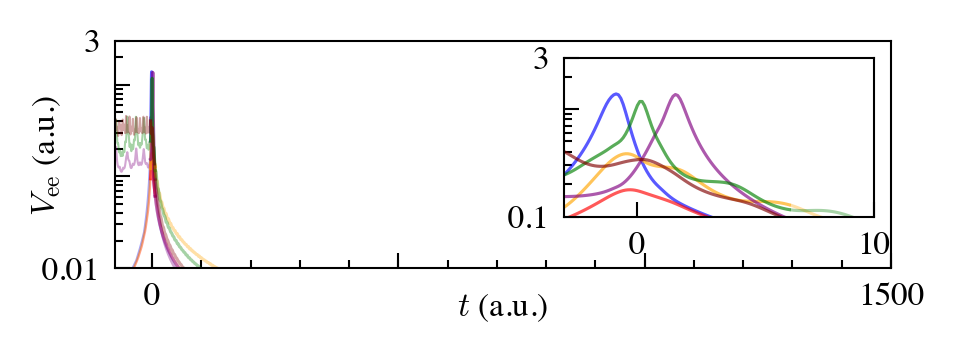}
        \caption{\label{ts1rectangle}}
    \end{subfigure}
    \caption{\textit{TS1} mechanism at $\epsilon_{0} = 95$ eV (see Sec.~\ref{sec:format_mechanisms} for details). }
    \label{fig:ts1}
\end{figure}

A typical \textit{TS1} mechanism is depicted in Fig.~\ref{fig:ts1}. In Fig.~\ref{ts1rightcircle}, the impact electron (black line) collides with and ionizes only one valence shell electron (blue (dark gray) line), and leaves the atom. This is the first step in the mechanism. Then, the blue (dark gray) valence shell electron, on its way out of the atom, collides with the other valence shell electron (red (medium gray) line), and both leave the atom. This is the second step in the mechanism. Figure~\ref{ts1rectangle} shows the timescale of the electron-electron interactions during \textit{TS1} processes. In the inset in Fig.~\ref{ts1rectangle}, we observe that the first peak in the electron-electron potential in blue corresponds to the first collision between the impact electron and the blue (dark gray) valence shell electron. The second peak in the electron-electron potential in green corresponds to the rapid transit of the blue (dark gray) valence shell electron through the inner shell region. The third peak in the electron-electron potential in purple corresponds to the last collision between the two valence shell electrons.
\par
A schematic diagram of \textit{TS1} mechanisms is depicted in the top left corner of Fig.~\ref{ts1leftcircle}. In \textit{TS1} mechanisms, the impact electron provides energy, which is greater than the sum of the first two ionization potentials of Mg, to one valence shell electron. This sets that (in this case the blue (dark gray)) valence shell electron on a trajectory towards ionization. In its transits through the inner shell region, the blue (dark gray) valence shell electron provides energy to the inner shell electron, but keeps enough energy to remain on its path towards ionization. Then, the collision between the two valence shell electrons leads to a redistribution of energy carried by the two electrons, such that the red (medium gray) valence shell electron is set on a trajectory towards ionization, while the other blue (dark gray) valence shell electron is reset onto a new trajectory towards ionization.

\subsubsection{TS2}

\begin{figure}[pt!]
\centering
    \begin{subfigure}{0.49\columnwidth}
        \includegraphics[trim = {0 0 1.64in 0}, clip]{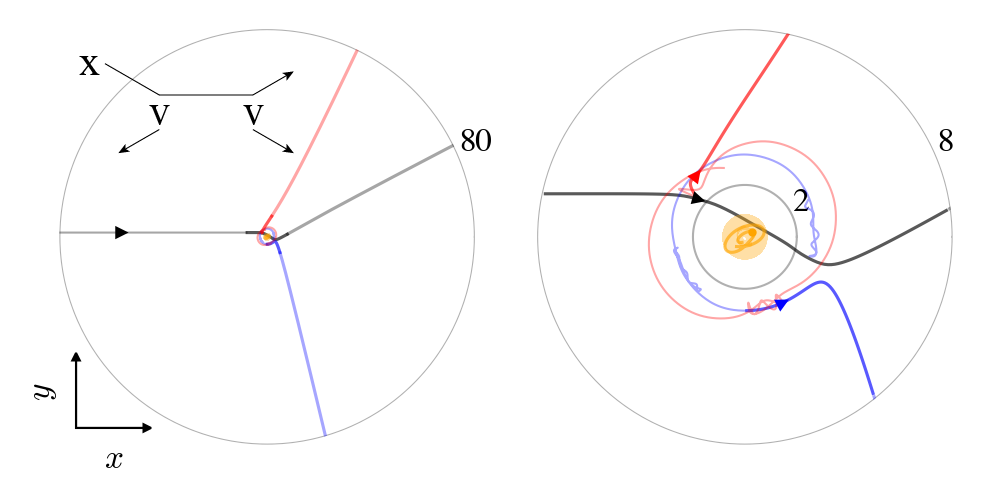}
        \caption{\label{ts2leftcircle}}
    \end{subfigure}
    \begin{subfigure}{0.49\columnwidth}
        \includegraphics[trim = {1.65in 0 0 0}, clip]{figure7ab.png}
        \caption{\label{ts2rightcircle}}
    \end{subfigure}
    \begin{subfigure}{\columnwidth}
        \includegraphics{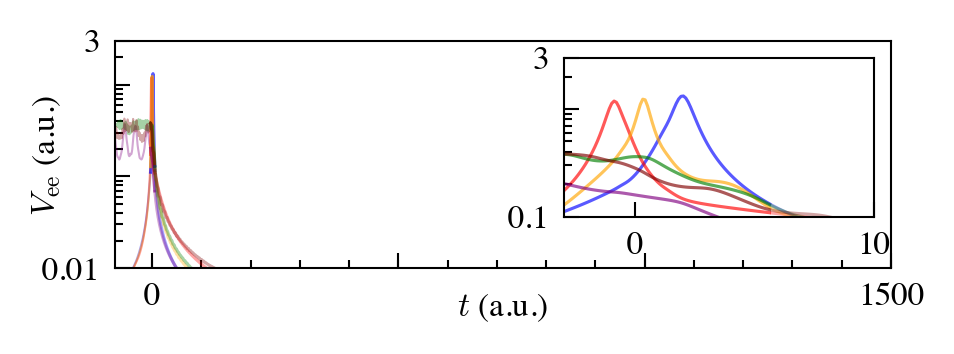}
        \caption{\label{ts2rectangle}}
    \end{subfigure}
    \caption{\textit{TS2} mechanism at $\epsilon_{0} = 95$~eV (see Sec.~\ref{sec:format_mechanisms} for details). }
    \label{fig:ts2}
\end{figure}

A typical \textit{TS2} mechanism is depicted in Fig.~\ref{fig:ts2}. In Fig.~\ref{ts2rightcircle}, the impact electron (black line) collides sequentially with and ionizes both valence shell electrons (blue (dark gray) and red (medium gray) lines) and leaves the atom. These are the two steps in the mechanism. Just as \textit{TS1} mechanisms, \textit{TS2} mechanisms occur on short time scales (typically, a few atomic units). In Fig.~\ref{ts2rectangle}, the first peak in the electron-electron potentials in red corresponds to the first collision between the impact electron and the red (medium gray) valence shell electron. The second peak in the electron-electron potential in yellow corresponds to the transit of the impact electron through the inner shell region. The third peak in the electron-electron potential in blue corresponds to the second collision between the impact electron and the other blue (dark gray) valence shell electron.
\par 
In the top left corner of Fig.~\ref{ts2leftcircle}, a schematic diagram of \textit{TS2} mechanisms is depicted. In \textit{TS2} mechanisms, the impact electron provides energy sequentially to the two valence shell electrons via collisions, which cause the valence shell electrons to ionize.

\subsection{Delay Mechanism \label{sec:delay_mechanisms}}

\textit{Delay} mechanisms are characterized by three features; first, the mechanisms occur on long time scales (typically, few tens of atomic units); second, the mechanisms do not involve any capture in the inner shell; and third, all valence shell electrons ionize. There are two types of \textit{Delay} mechanisms, \textit{D1} and \textit{D0}, that are represented in Fig.~\ref{fig:d1} and in Fig.~\ref{fig:d0}, respectively. In the nomenclature of the mechanisms, `D' refers to \textit{Delay} mechanisms. `1' or `0' refers to the number of electrons that are ionized before the threshold time $t = t_{\textrm{Th}}$, or, in other words, the number of electrons which are ionized on a short timescale. 

\subsubsection{D1}

\begin{figure}[pt!]
\centering
    \begin{subfigure}{0.49\columnwidth}
        \includegraphics[trim = {0 0 1.64in 0}, clip]{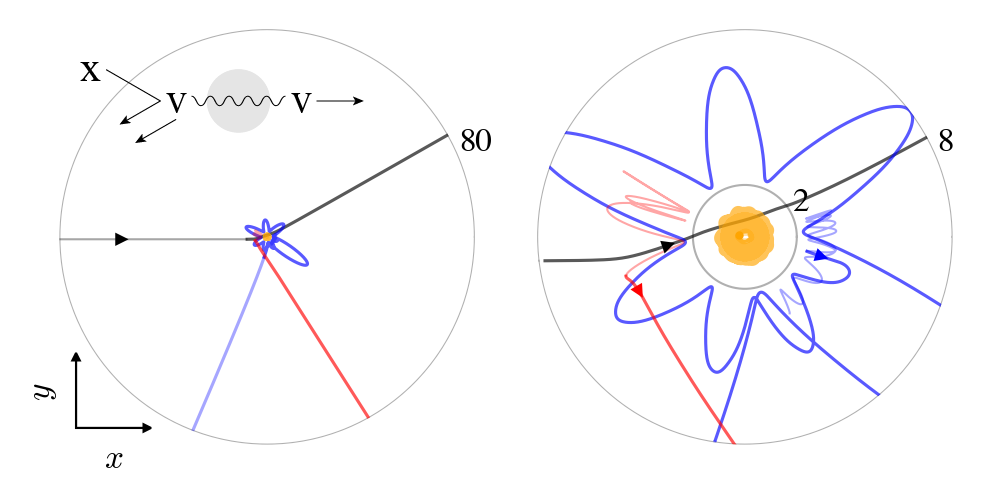}
        \caption{\label{d1leftcircle}}
    \end{subfigure}
    \begin{subfigure}{0.49\columnwidth}
        \includegraphics[trim = {1.65in 0 0 0}, clip]{figure8ab.png}
        \caption{\label{d1rightcircle}}
    \end{subfigure}
    \begin{subfigure}{\columnwidth}
        \includegraphics{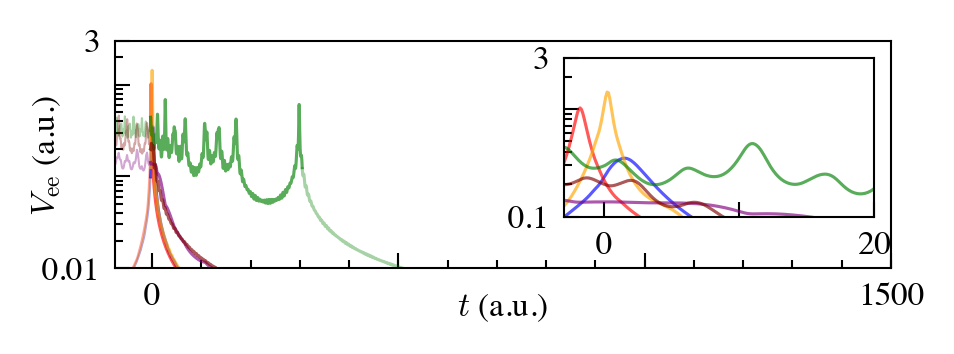}
        \caption{\label{d1rectangle}}
    \end{subfigure}
    \caption{\textit{D1} mechanism at $\epsilon_{0} = 95$~eV (see Sec.~\ref{sec:format_mechanisms} for details). }
    \label{fig:d1}
\end{figure}

A typical \textit{D1} mechanism is depicted in Fig.~\ref{fig:d1}. In Fig.~\ref{d1rightcircle}, the impact electron (black line) first collides with a valence shell electron (red (medium gray) line). Then, the impact electron transits through the inner shell region and leaves the atom. In the inset in Fig.~\ref{d1rectangle}, the electron-electron potential between the impact electron and the red (medium gray) valence shell electron in red peaks first. It indicates that the first collision occurs between the impact electron and the red (medium gray) valence shell electron. This peak is followed by the peak in the electron-electron potential between the impact electron and the inner shell electron in yellow. It indicates the rapid transit of the impact electron through the inner shell region. The steep decrease in the red, blue, yellow, brown and purple (all but green) electron-electron potentials implies that the impact electron and the red (medium gray) valence shell electron leave the atom. Thus, the atom is reduced to a two-electron ion. Several peaks in the green electron-electron potential between the blue (dark gray) valence shell electron and the inner shell electron imply that the blue (dark gray) valence shell electron collides several times with the inner shell electron, with rather large excursions from the ionic core. The blue (dark gray) valence shell electron eventually ionizes.
\par
The electron-electron potential in green (the one that prolongs) between the blue (dark gray) valence shell electron and the inner shell electron after $t = t_{\textrm{Th}}$ highlights the three features of the \textit{D1} mechanism; first, the initial collision between the impact electron and the red valence shell electron ionizes the later before $t = t_{\textrm{Th}}$; second, the impact electron excites the inner shell electron, when it transits rapidly through the inner shell region. Therefore, the two-electron ion that forms during the mechanism has net positive energy, which is redistributed between the inner shell electron and the blue (dark gray) valence shell electron through collisions; And third, the \textit{D1} mechanism relies on a strong electron-electron correlation between the inner shell electron and valence shell electron, and, thus, without the existence of an inner shell electron, the \textit{D1} mechanism cannot occur. The schematic diagram in the top left of Fig.~\ref{d1leftcircle} summarizes the important features of the mechanism; one valence shell electron ionizes before $t = t_{\textrm{Th}}$, while the other ionizes with a delay because of the electron-electron correlation with the excited inner shell electron.

\subsubsection{D0}

\begin{figure}[pt!]
\centering
    \begin{subfigure}{0.49\columnwidth}
        \includegraphics[trim = {0 0 1.64in 0}, clip]{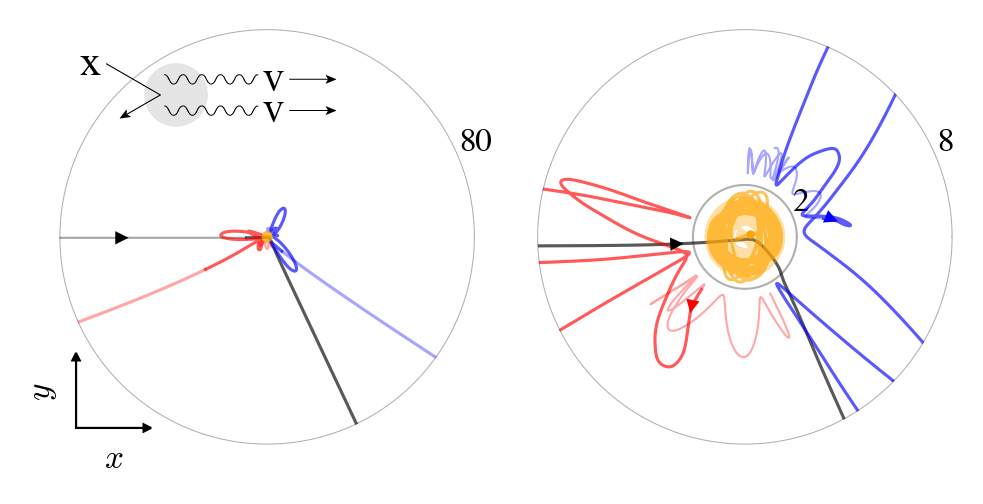}
        \caption{\label{d0leftcircle}}
    \end{subfigure}
    \begin{subfigure}{0.49\columnwidth}
        \includegraphics[trim = {1.65in 0 0 0}, clip]{figure9ab.png}
        \caption{\label{d0rightcircle}}
    \end{subfigure}
    \begin{subfigure}{\columnwidth}
        \includegraphics{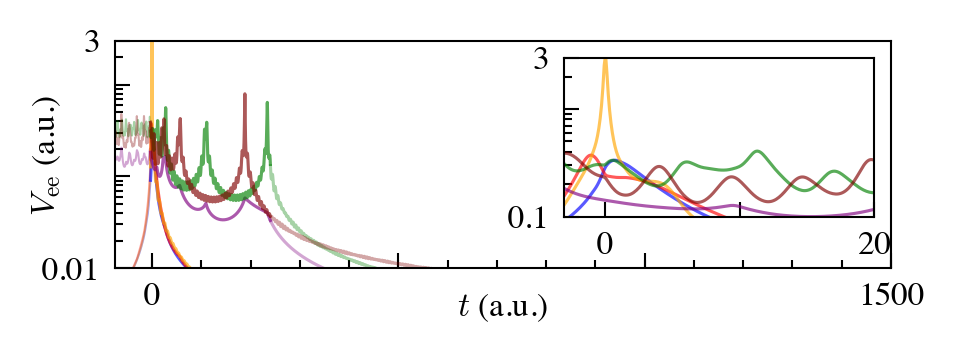}
        \caption{\label{d0rectangle}}
    \end{subfigure}
    \caption{\textit{D0} mechanism at $\epsilon_{0} = 95$~eV (see Sec.~\ref{sec:format_mechanisms} for details). }
    \label{fig:d0}
\end{figure}

A typical \textit{D0} mechanism is depicted in Fig.~\ref{fig:d0}. In Fig.~\ref{d0rightcircle}, the impact electron (black line) first enters the inner shell region and collides with the inner shell electron. As a result, the impact electron gets deflected and leaves the atom. In Fig.~\ref{d0rectangle}, this collision appears as a peak in the yellow electron-electron potential between the impact electron and inner shell electron. The ionized impact electron leaves behind an excited three-electron atom. The two valence shell electrons in the atom undergo multiple collisions with the inner shell electron, potentially with large excursions from the ionic core. In Fig.~\ref{d0rectangle}, these collisions appear as successive peaks in the brown electron-electron potential (second highest at $t = 100$~a.u.)\ between red (medium gray) valence shell electron and inner shell electron, and green electron-electron potential (first highest at $t = 100$~a.u.)\ between the blue (dark gray) valence shell electron and the inner shell electron. Through collisions inside the excited atom, the red (medium gray) valence shell electron ionizes first and leaves behind an excited two electron ion. Eventually, the blue (dark gray) valence shell electron also ionizes.
\par
The electron-electron potential in brown (second highest at $t = 100$~a.u.)\ between the red (medium gray) valence shell electron and the inner shell electron, and the electron-electron potential in green (first highest at $t = 100$~a.u.)\ between the blue (dark gray) valence shell electron and the inner shell electron, beyond $t = t_{\textrm{Th}}$, bring the three significant features of the \textit{D0} mechanism to light; first, none of the electrons ionizes from the atom before $t = t_{\textrm{Th}}$, but the atom is left in an excited state, that is, with net positive energy; second, the ionization occurs because of a redistribution of the net positive energy, through multiple collisions among the valence shell and inner shell electrons; and third, the magnitude of the electron-electron potentials of the valence shell electrons with the inner shell electron (brown and green electron-electron potentials) is greater than electron-electron potential between the two valence shell electrons in purple (third highest at $t = 100$~a.u.), which implies that the mechanism mainly relies on the electron-electron correlations of the valence shell electrons with the inner shell electron. Therefore, just like \textit{D1}, \textit{D0} requires the existence an inner shell electron to take place. Moreover, \textit{D0} involves a complex (chaotic) choreography between the three electrons of the target. 

\subsection{Ionized Inner Shell Mechanisms \label{sec:inner_shell_mechanisms}}

\textit{Ionized Inner Shell} mechanisms involve the ionization of the inner shell electron. There are two types of \textit{Ionized Inner Shell} mechanisms: \textit{Direct*} and \textit{Delay*} mechanisms, which are in turn composed of \textit{TS1*} and \textit{TS2*} mechanisms, and \textit{D1*} mechanisms, respectively. The electron dynamics in \textit{TS1*}, \textit{TS2*} and \textit{D1*} mechanisms are similar to the electron dynamics in \textit{TS1}, \textit{TS2} and \textit{D1} mechanisms, respectively, with the exception that the inner shell electron is ionized in \textit{TS1*}, \textit{TS2*} and \textit{D1*} mechanisms. We use an asterisk in the nomenclature to differentiate these mechanisms from the ones that do not (or very weakly) involve the inner shell electron.
\par
The contribution of \textit{TS1*}, \textit{TS2*} and \textit{D1*} mechanisms to the probability of \textit{Ionized Inner Shell} mechanisms is depicted in Fig.~\ref{fig:detailedcrosssectionlowerpanel}. We notice that we have not mentioned any \textit{D0*} mechanisms because they occur with negligible probability in our calculations. This also explains why the probability of \textit{D1*} mechanisms overlaps the probability of \textit{Delay*} mechanisms in Fig.~\ref{fig:detailedcrosssectionlowerpanel}. Figure~\ref{fig:detailedcrosssectionupperpanel} illustrates that the \textit{Inner Shell Ionized} mechanisms begin to occur at about $\epsilon_0 = 80$~eV, and are dominant over \textit{TS1} and \textit{D0} mechanisms at $\epsilon_0 = 125$~eV.

\subsubsection{TS1*, Direct*}

\begin{figure}[pt!]
\centering
    \begin{subfigure}{0.49\columnwidth}
        \includegraphics[trim = {0 0 1.64in 0}, clip]{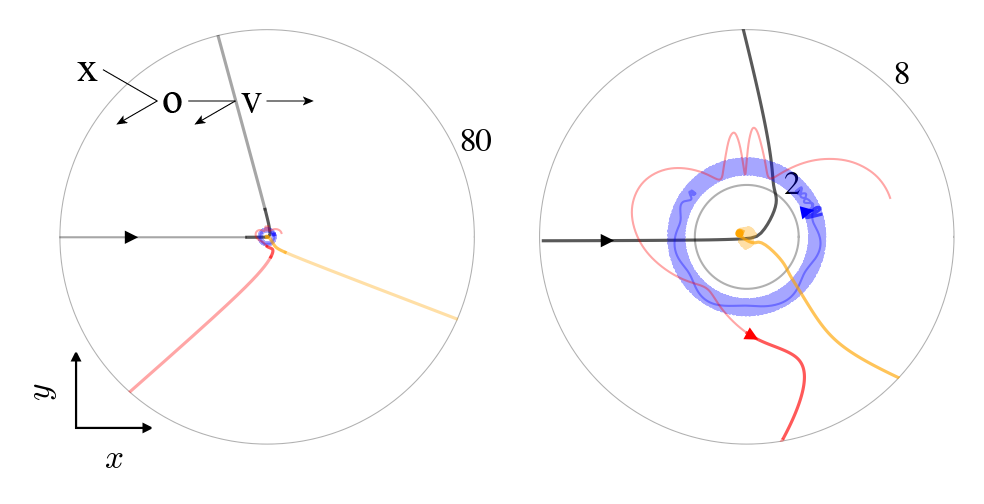}
        \caption{\label{ts1starleftcircle}}
    \end{subfigure}
    \begin{subfigure}{0.49\columnwidth}
        \includegraphics[trim = {1.65in 0 0 0}, clip]{figure10ab.png}
        \caption{\label{ts1starrightcircle}}
    \end{subfigure}
    \begin{subfigure}{\columnwidth}
        \includegraphics{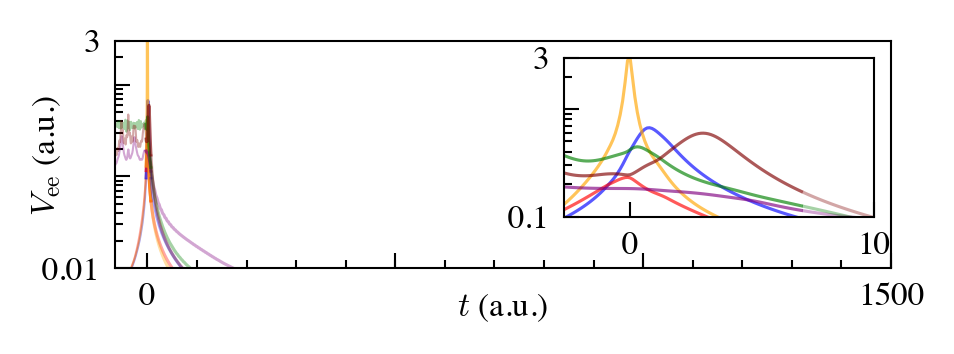}
        \caption{\label{ts1starrectangle}}
    \end{subfigure}
    \caption{\textit{TS1*} mechanism at $\epsilon_{0} = 125$~eV (see Sec.~\ref{sec:format_mechanisms} for details). }
    \label{fig:ts1star}
\end{figure}

A typical \textit{TS1*} mechanism is depicted in Fig.~\ref{fig:ts1star}. From the zoomed-out view of the electron trajectories in Fig.~\ref{ts1starleftcircle}, we see that this is an \textit{Ionized Inner Shell} trajectory, because it leads to the ionization of the inner shell electron and a valence shell electron, trajectories, which are traced in yellow (light gray) and red (medium gray) respectively. Figure~\ref{ts1starrightcircle} shows that the impact electron travels into the inner shell region, undergoes a strong collision with the inner shell electron and the blue (dark gray) valence shell electron, and deflects out of the atom. As a result of its collision with the impact electron, the inner shell electron escapes the inner shell region, and, on its trajectory towards ionization, collides with the red (medium) valence shell electron. That collision sets the red (medium gray) valence shell electron also on a path towards ionization, and the collision resets the inner shell electron's trajectory towards ionization. The schematic diagram in the top left corner of Fig.~\ref{ts1starleftcircle} summarizes the main features of the mechanism.

\subsubsection{TS2*, Direct*}

\begin{figure}[pt!]
\centering
    \begin{subfigure}{0.49\columnwidth}
        \includegraphics[trim = {0 0 1.64in 0}, clip]{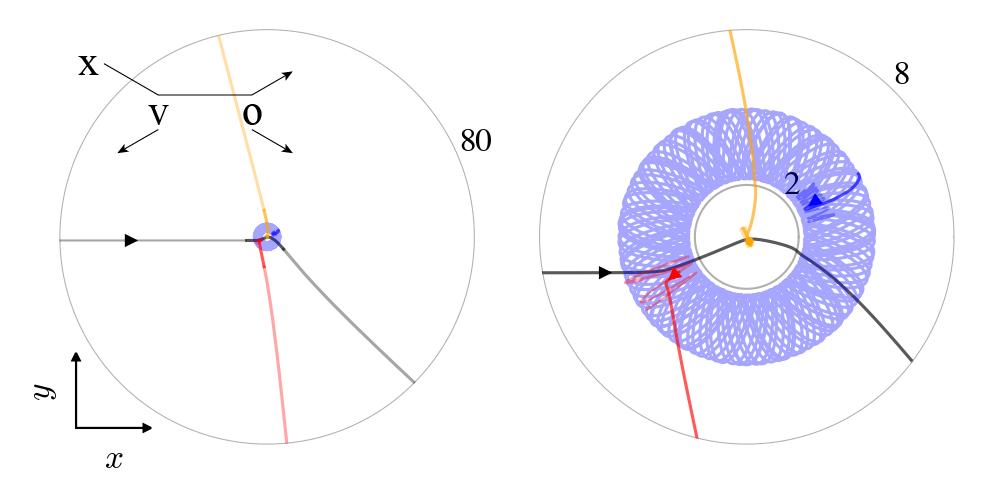}
        \caption{\label{ts2starleftcircle}}
    \end{subfigure}
    \begin{subfigure}{0.49\columnwidth}
        \includegraphics[trim = {1.65in 0 0 0}, clip]{figure11ab.png}
        \caption{\label{ts2starrightcircle}}
    \end{subfigure}
    \begin{subfigure}{\columnwidth}
        \includegraphics{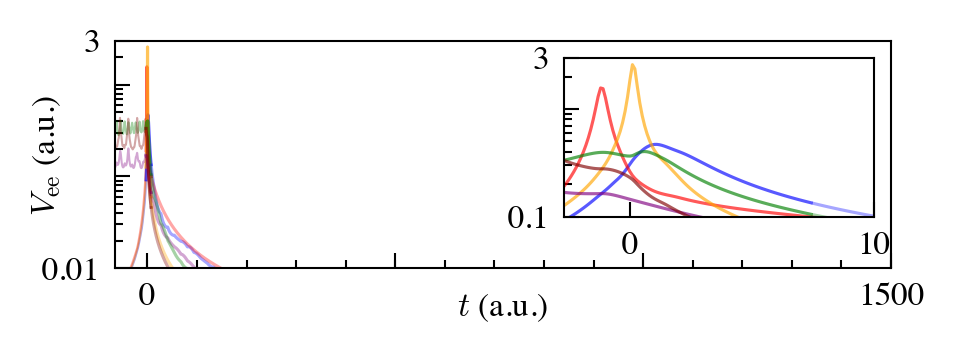}
        \caption{\label{ts2starrectangle}}
    \end{subfigure}
    \caption{\textit{TS2*} mechanism at $\epsilon_{0} = 125$~eV (see Sec.~\ref{sec:format_mechanisms} for details). }
    \label{fig:ts2star}
\end{figure}

A typical \textit{TS2*} mechanism is depicted in Fig.~\ref{fig:ts2star}. Again, the zoomed out view of the electron trajectories in Fig.~\ref{ts2starleftcircle} illustrates that the trajectory corresponds to an \textit{Ionized Inner Shell} mechanism, because it leads to the ionization of the inner shell electron and a valence shell electron, trajectories, which are traced in yellow (light gray) and red (medium gray), respectively. Figure~\ref{ts2starrightcircle} shows that the impact electron first collides with the red (medium gray) valence shell electron before traveling into the inner shell region. In the inner shell region, the impact electron collides with the inner shell electron. Next it collides with the blue (dark gray) valence shell electron, before deflecting out of the atom. As a result of their collisions with the impact electron, the inner shell electron and red (medium gray) valence shell electron ionize. In Fig.~\ref{ts2starrectangle}, the collision between the impact electron and the inner shell electron appears as a peak in the electron-electron potential in yellow between the impact electron and the inner shell electron. The sequence and duration of the events in this mechanism are schematically depicted in the diagram in the top left corner of Fig.~\ref{ts2starleftcircle}.

\subsubsection{D1*, Delay*}

\begin{figure}[pt!]
\centering
    \begin{subfigure}{0.49\columnwidth}
        \includegraphics[trim = {0 0 1.64in 0}, clip]{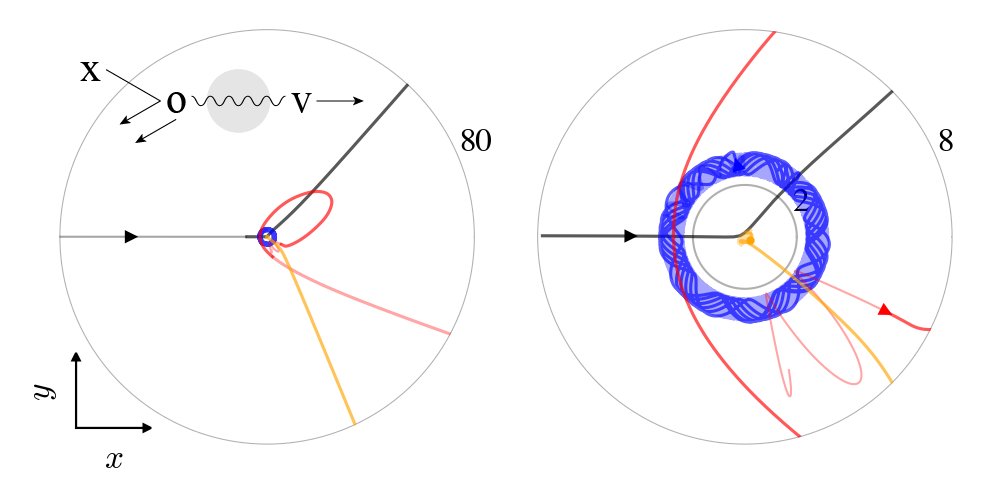}
        \caption{\label{d1starleftcircle}}
    \end{subfigure}
    \begin{subfigure}{0.49\columnwidth}
        \includegraphics[trim = {1.65in 0 0 0}, clip]{figure12ab.png}
        \caption{\label{d1starrightcircle}}
    \end{subfigure}
    \begin{subfigure}{\columnwidth}
        \includegraphics{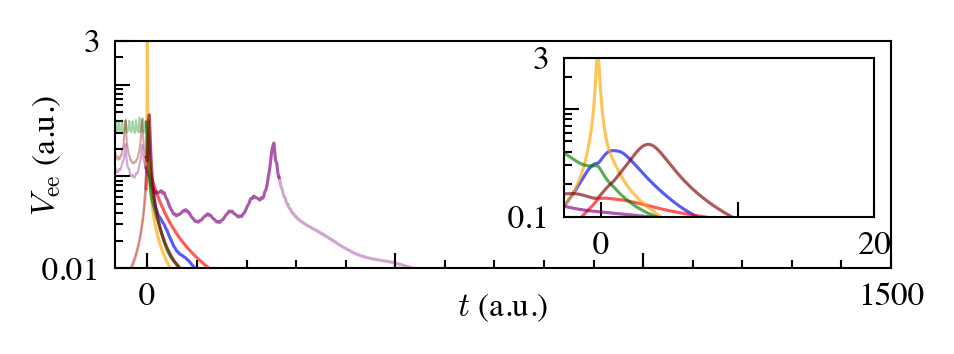}
        \caption{\label{d1starrectangle}}
    \end{subfigure}
    \caption{\textit{D1*} mechanism at $\epsilon_{0} = 125$~eV (see Sec.~\ref{sec:format_mechanisms} for details). }
    \label{fig:d1star}
\end{figure}

A typical \textit{D1*} mechanism is depicted in Fig.~\ref{fig:d1star}. From the zoomed-out view of the electron trajectories in Fig.~\ref{d1starleftcircle}, we conclude that trajectory depicts an \textit{Ionized Inner Shell} mechanism, because it leads to the ionization of the inner shell electron and a valence shell electron, which are traced in yellow (light gray) and red (medium gray) respectively. As inferred from Fig.~\ref{d1starrightcircle}, the impact electron travels into the inner shell region, collides with the inner shell electron and leaves the atom. As a result of that collision, the inner shell electron (yellow (light gray) line) is set on trajectory towards ionization. On its path, the inner shell electron collides with the red (medium gray) valence shell electron and then leaves the atom. A two-electron ion results, in which both the electrons are valence shell electrons. When the red (medium gray) valence shell electron passes near the 2~a.u.\ boundary, the energy is redistributed between the two valence shell electrons, with a greater share belonging to the red (medium gray) valence shell electron. Therefore, the red (medium gray) valence shell electron ionizes. Information about the main actors in the collisions and the time scales involved in these collisions is contained in the schematic diagram in the top left corner of Fig.~\ref{d1starleftcircle}. 
As a side note, since this mechanism involves a direct ionization of the inner shell electron, followed by a relaxation towards a stable state by a subsequent ionization, this is a classical equivalent of the Auger effect. From our analysis, we see that this is a very probable mechanism at impact energies above 100~eV, but it is not the dominant mechanism, since it appears that direct processes (as the combination of the two direct processes \textit{TS1*} and \textit{TS2*}) are at least as present as \textit{D1*}.

\subsection{Selecting mechanisms by varying the impact parameter $y_{0, \textrm{in}}$}

\begin{figure*}
\centering
    \begin{subfigure}{0.32\textwidth}
        \includegraphics{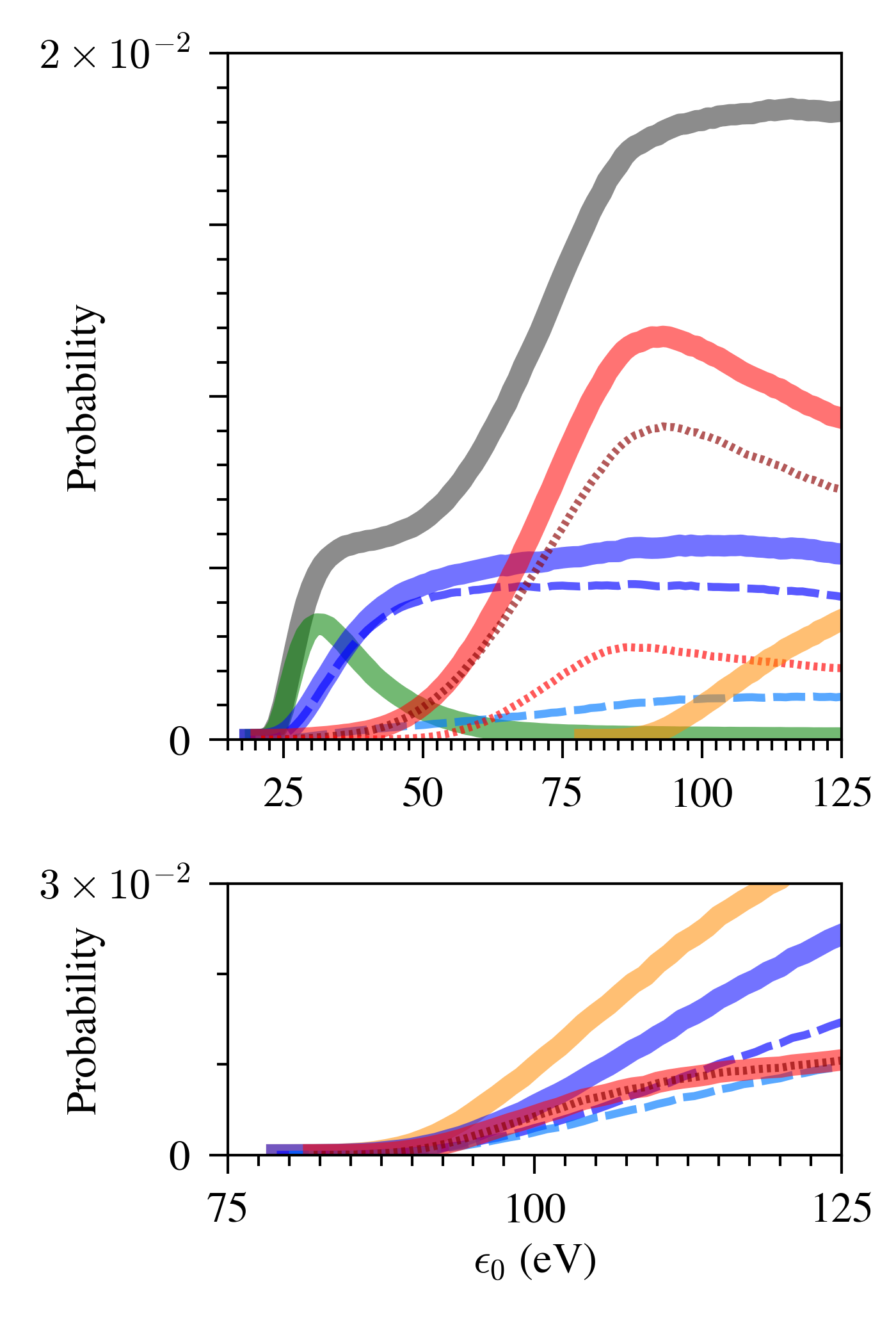}
        \caption{\label{discussion1a}Impact parameters in $\left[0, 2\right)$~a.u.}
    \end{subfigure}
    \begin{subfigure}{0.32\textwidth}
        \includegraphics{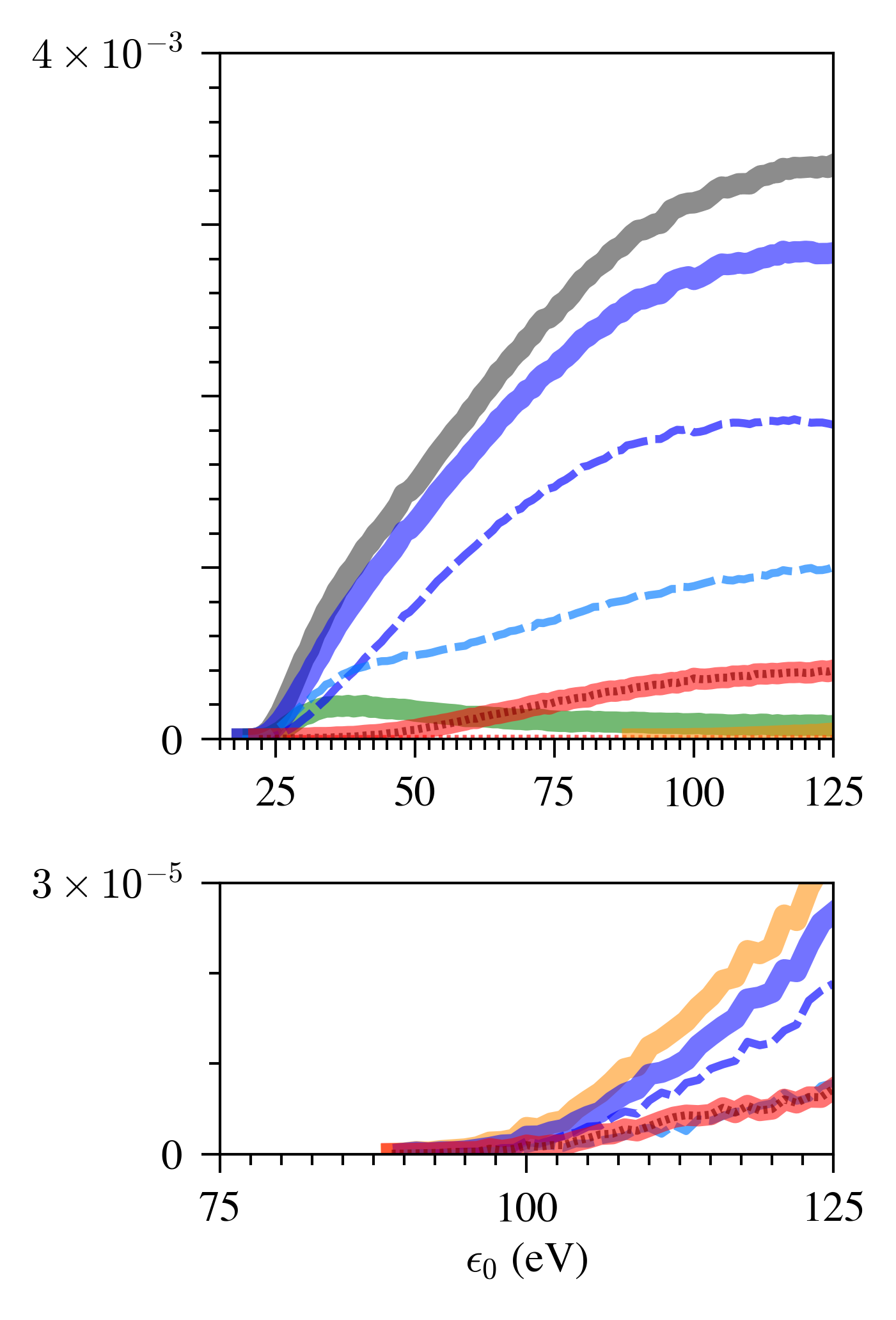}
        \caption{\label{discussion1b}Impact parameters in $\left[2, 3\right]$~a.u.}
    \end{subfigure}
    \begin{subfigure}{0.32\textwidth}
        \includegraphics{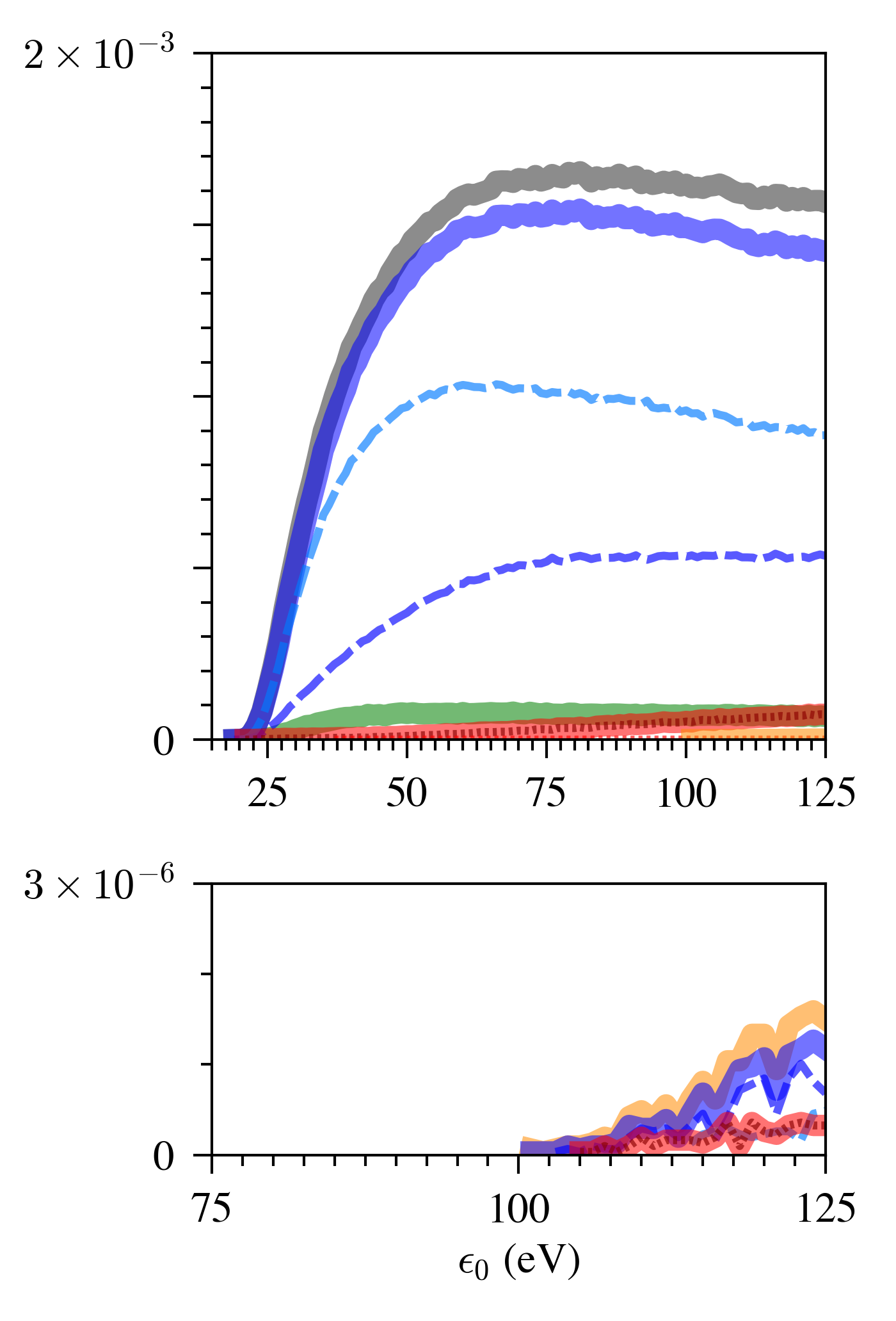}
        \subcaption{\label{discussion1c}Impact parameters in $\left(3, 5\right]$~a.u.}
    \end{subfigure}
    \caption{Double ionization probability versus impact energy $\epsilon_0$ for three different sets of impact parameters: (\subref{discussion1a})  $ y_{0, \textrm{in}} \in \left[0, 2\right)$~a.u., (\subref{discussion1b}) $y_{0, \textrm{in}} \in \left[2, 3\right]$~a.u., and (\subref{discussion1c})  $y_{0, \textrm{in}} \in \left(3, 5\right]$~a.u.\ We consider the entire domain of impact energies, $\epsilon_0 \in \left[15, 125\right]$~eV in the upper sub-panels and a magnified domain, $\epsilon_0\in \left[75, 125\right]$~eV in the corresponding lower sub-panels. Notice that the ranges in the vertical axes are different for the different sub-panels. The color code of the different mechanisms is the same as in Fig.~\ref{fig:detailedcrosssection}.}
    \label{fig:discussion1}
\end{figure*}

\textit{Inner Shell Capture}, \textit{D1}, \textit{D0}, \textit{TS1*}, \textit{TS2*} and \textit{D1*} mechanisms rely on electron-electron correlations with the inner shell electron. These correlations are pronounced because the impact electron excites or ionizes the inner shell electron when it transits through the inner shell region in these mechanisms. Therefore, by increasing the impact parameter, $y_{0, \textrm{in}}$, the probabilities of occurrences of the six mechanisms that rely on electron-electron correlations with the inner shell electron might decrease. We decompose the probability curves in Fig.~\ref{fig:discussion1} for three ranges of impact parameters. We show that decomposition of probabilities for each of the three sets of impact parameters in the three panels of Fig.~\ref{fig:discussion1}, where the panels Figs.~\ref{discussion1a}, \ref{discussion1b} and \ref{discussion1c} represent the probabilities for impact parameter $y_{0, \textrm{in}}$ in $[0, 2)$~a.u., $[2, 3]$~a.u.\ and $(3, 5]$~a.u., respectively. Each panel in Fig.~\ref{fig:discussion1} follows the legend of Fig.~\ref{fig:detailedcrosssection}.
\par
The upper panels in Fig.~\ref{fig:discussion1} show that the probabilities of \textit{Inner Shell Capture}, \textit{D1} and \textit{D0} mechanisms are progressively suppressed by one order of magnitude from the leftmost to rightmost panel. In other words, the probabilities of \textit{Inner Shell Capture}, \textit{D1} and \textit{D0} mechanisms are significantly reduced when the impact parameter is increased from $[0, 2)$~a.u.\ to $(3, 5]$~a.u.\ Moreover, the lower panels in Fig.~\ref{fig:discussion1} show that the probabilities of \textit{TS1*}, \textit{TS2*} and \textit{D1*} mechanisms are suppressed by four orders of magnitude from the leftmost panel to the rightmost one. This indicates that the probabilities of \textit{TS1*}, \textit{TS2*} and \textit{D1*} mechanisms are drastically suppressed when the impact parameter increases from $[0, 2)$~a.u.\ to $(3, 5]$~a.u.
\par
According to Fig.~\ref{discussion1c}, our model predicts that \textit{Direct} mechanisms dominate the probability of double ionization of Mg by electron impact for large impact parameters. Furthermore, our model predicts that \textit{TS1} mechanisms dominate all other mechanisms in large impact parameter scenarios. The impact electron misses the inner shell when it arrives at large impact parameters. All electron-electron correlations with the inner shell electron remain unchanged, and electron-electron correlations with the valence shell electrons drive the ionization mechanisms. Furthermore, the two valence shell electrons are likely to occupy opposite sites in the atom because of electron-electron repulsion. And, when the impact electron hits the atom at large impact parameters, the impact electron is likely to collide with only one of the valence shell electron, closest to it. Therefore, \textit{TS1} \textit{Direct} mechanisms become dominant. By the same logic, if the impact electron arrives at small impact parameters, then electron-electron correlations with inner shell electron become significant and mechanisms other than the \textit{Direct} mechanisms become dominant. Also, the impact electron is likely to collide with both valence shell electrons in small impact parameter scenarios. Consequently, \textit{TS2} mechanism becomes the dominant \textit{Direct} mechanism for small impact parameters, which is observed in Figs.~\ref{discussion1a} and~\ref{discussion1b}.
\par
In general, different mechanisms are dominant at different impact energies as depicted in Fig.~\ref{fig:detailedcrosssection}. Moreover, we find that different mechanisms can be made dominant by appropriately selecting the impact parameters. Therefore, the impact energy, $\epsilon_0$, and the impact parameter, $y_{0, \textrm{in}}$, may be tuned to selectively trigger or suppress certain double ionization mechanisms.

\section{Discussions \label{sec:discussion}}

\subsection{Comparison with a two electron model \label{sec:two_electron_discussion}}

Up to now we have described our model and its two salient features: first, a successful qualitative reproduction of the probability of double ionization of Mg by electron impact, and second, the prediction about the underlying double ionization mechanisms. In this section, we compare our model and its implications with the two bound electron model of Mg in Ref.~\cite{Dubois2017}. This comparison will highlight the importance of considering an inner shell electron and the importance of fully accounting for all electron-electron correlations with the inner shell electron.
\par
Ref.~\cite{Dubois2017} considers a two-active electron model to study the single and double ionization of Mg by electron impact. Ref.~\cite{Dubois2017} also uses a purely classical mechanical approach to describe the double ionization processes and all electron-electron correlations taken into account. However, the model in Ref.~\cite{Dubois2017} differs from the model in this paper in two ways; first, the atom consists of an ionic core and two valence shell electrons only, and second, a soft-Coulomb potential without effective charge governs the electron-core potential. Appropriately adjusting the soft-Coulomb parameters $a$ and $b$ alone guarantees no self ionization and non-empty ground states in the two-electron model in Ref.~\cite{Dubois2017}. Our model considers an inner shell electron in addition to an ionic core and two valence shell electrons, and the existence of the inner shell electron necessitates an effective charge in the soft-Coulomb electron-core potential in our model.
\par
The two-electron model in Ref.~\cite{Dubois2017} explains the first increase in the double ionization probability and predicts that only \textit{TS1} and \textit{TS2} \textit{Direct} mechanisms contribute to the double ionization probability. It also predicts that \textit{TS2} mechanisms are dominant at small impact parameters. The model reproduces the double ionization probability of Mg by electron impact for $20 < \epsilon_0 < 50$~eV, but it fails to explain the knee in the double ionization probability. This indicates that the knee in the probability of double ionization of Mg by electron impact must be a consequence of electron-electron correlations with the inner shell electron(s). Figure~\ref{fig:detailedcrosssectionupperpanel}, which shows the contribution of the \textit{Delay} and the \textit{Ionized Inner Shell} mechanisms to the increase in the double ionization probability for $\epsilon_0 > 50$~eV, substantiates this claim.
\par
Also as a result of the differences between a two-electron and a three-electron model, the former can only predict \textit{TS1} and \textit{TS2} \textit{Direct} mechanisms, which require electron-electron correlations among an impact electron and two valence shell electrons only. Our model predicts six more mechanisms, other than \textit{TS1} and \textit{TS2}, because our model accounts for all electron-electron correlations among an impact electron, two valence shell electrons and an inner shell electron.
\par
Another consequence of the differences between a two-electron and three-electron models is the nature of mechanisms which underlie the probability of double ionization for $20 < \epsilon_0 < 50$~eV. Ref.~\cite{Dubois2017} finds that only the \textit{TS1} and \textit{TS2} mechanisms contribute to the double ionization probability in that domain of impact energies. However, we find that \textit{Inner Shell Capture} mechanisms also contribute substantially to the total double ionization probability in the same domain of impact energies. For large impact parameters and in the intermediate range of impact energies, we find \textit{Direct} mechanisms dominate over \textit{Delay} mechanisms, which agrees with Ref.~\cite{Dubois2017}.
\par
Altogether, the models in Ref.~\cite{Dubois2017} and in this paper are different with respect to the number of components in the models and the structure of the soft-Coulomb electron-core potentials used in the models. The differences result in different sets of possible electron-electron correlations. The different sets of electron-electron correlations, in turn, affect the possible double ionization mechanisms, their probabilities and the probability of double ionization of Mg by electron impact. Some agreement can be found between these models in the intermediate range of impact energies and relatively high impact parameters (in order to suppress the role of the inner shell electron). 

\subsection{Relevance of the model with respect to experimental results}

So far, the discussion focused on the double ionization of Mg by electron impact for $\epsilon_0 < 125$~eV. However, experiments on double ionization by electron impact are also conducted for $200 < \epsilon_0 < 5000$~eV. In order to check how our model compares with these experiments, we compute the probabilities of the eight double ionization mechanisms predicted by our model at $\epsilon_0 = 700$~eV. In what follows, we first explain the relevant experiments, and then link results from our model with experimental observations.
\par
In the experimental measurements reported in Ref.~\cite{LahmamBennani2001}, we observe an asymmetric differential cross-section in electron impact double ionization of Helium. In Ref. ~\cite{LahmamBennani2001}, it is suggested that this asymmetry implies the prominence of second or higher order mechanisms over first order mechanisms. We identify TS1 and TS1* mechanisms in our analysis as the first order mechanisms, and all other mechanisms in our analysis as second or higher order mechanisms. We refer to Refs.~\cite{Goetz2003, LahmamBennani2010, Casagrande2011} for a more detailed discussion on this issue. Moreover, in Ref.~\cite{Casagrande2011}, it is suggested that non-first order mechanisms, other than \textit{TS2}, may be responsible for the disagreement between their theoretical and experimental results. Hence, experiments on double ionization by electron impact suggest prominence of non-first order over first order double ionization mechanisms. So far such experiments probing the prominence of first and non-first order mechanisms have been conducted for Argon, Neon, molecular Nitrogen and Helium for impact energies approximately between 600 and 700~eV~\citep{LahmamBennani2001,LahmamBennani2010,Casagrande2011,Li2011,Li2012}.
\par
\begin{center}
\begin{figure}[pt!]
    \includegraphics[]{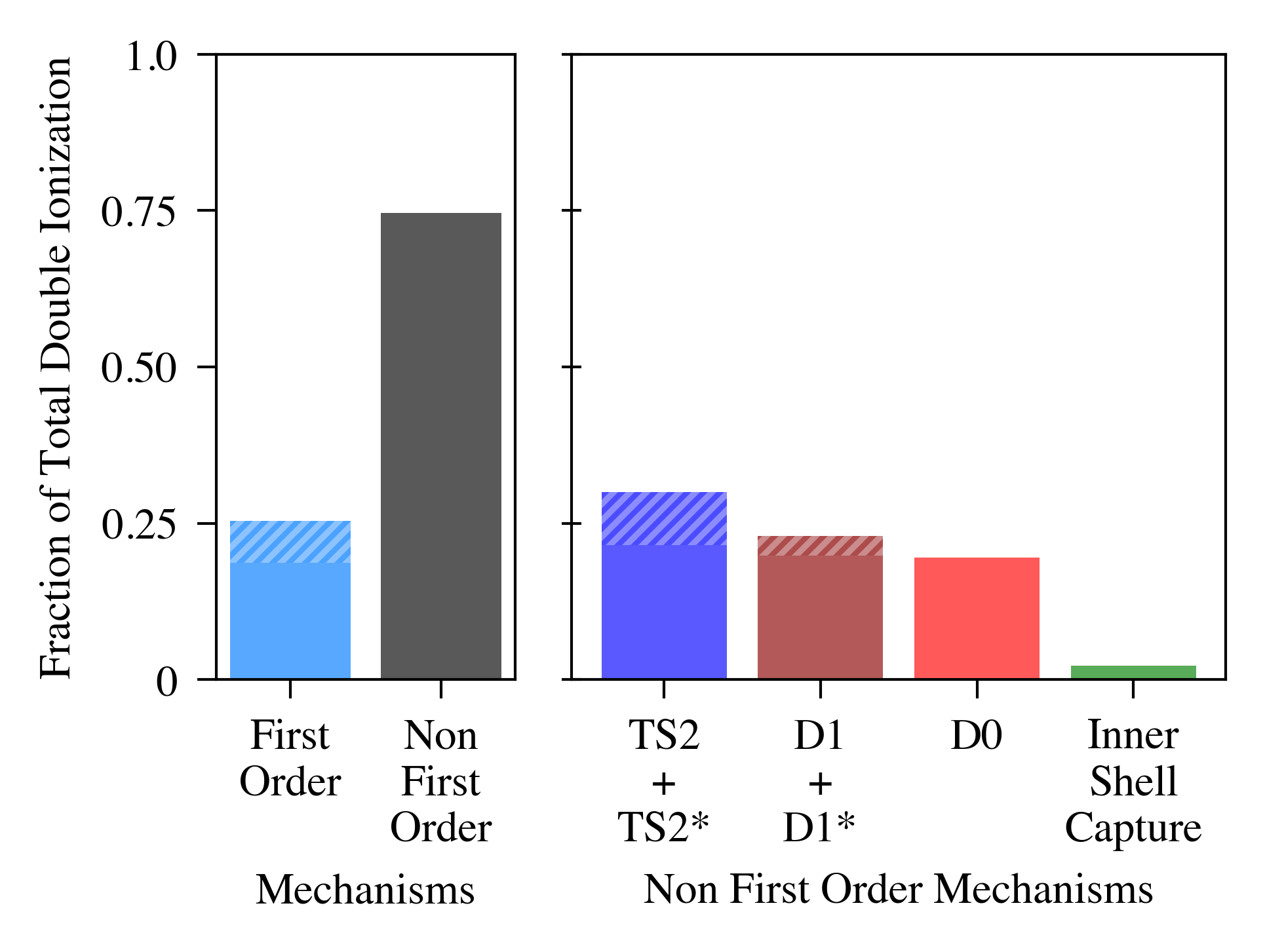}
    \caption{Relative prominence of double ionization mechanisms at $\epsilon_0 = 700$~eV. The first order mechanisms are the \textit{TS1} and \textit{TS1*} mechanisms. The hatched section in the bars represents the mechanisms whose denominations contain `*'.}
    \label{fig:discussion2}
\end{figure}
\end{center}

The probability of first and non-first order mechanisms at high impact energies of Hamiltonian~\eqref{eq:total_Hamiltonian}, for $\epsilon_0 = 700$~eV, is shown in the form of a bar graph in Fig.~\ref{fig:discussion2}. The vertical axis measures the contribution of each mechanism as a fraction of the total count of double ionization of Mg by electron impact. The left panel in Fig.~\ref{fig:discussion2} differentiates first order mechanisms, \textit{TS1} and \textit{TS1*}, represented by blue (gray) bars, from non-first order mechanisms, the remaining six mechanisms represented by the black bar. The non-first order mechanisms are broken up into constituent mechanisms in the right panel in Fig.~\ref{fig:discussion2}. The hatched sections in the bars represent the contribution from those mechanisms in which the inner shell electron is ionized, that is, from those mechanisms with `*' in their nomenclature. The left panel in Fig.~\ref{fig:discussion2} exhibits a clear agreement between results from our model and experimental observation, because our model predicts, as is observed in the experiments, that non-first order mechanisms are more probable than, and, thus, dominant over, first order mechanisms at 700~eV.

\section{Conclusion}

We have introduced a model to study the double ionization of Mg atom by electron impact. In this model, we consider two valence shell electrons, one inner shell electron and an ionic core. The electrons are bounded to the core using a soft-Coulomb electron-core potential with an effective charge. We find that the probability of double ionization of the Mg atom as a function of the initial energy of the impact electron calculated from our model is qualitatively similar to that obtained from experiments. Our model enables us to analyze electron trajectories of double ionization mechanisms, and based on visual inspection of the electron dynamics in trajectories, we find eight dominant mechanisms - \textit{Inner Shell Capture}, \textit{TS1}, \textit{TS2}, \textit{D1}, \textit{D0}, \textit{TS1*}, \textit{TS2*}, and \textit{D1*}. We group the eight mechanisms into four broad categories - \textit{Inner Shell Capture}, \textit{Direct}, \textit{Delay} and \textit{Ionized Inner Shell} mechanisms. We find that \textit{Delay} and the \textit{Ionized Inner Shell} mechanisms underlie the knee in the double ionization probability. We provide a detailed description of the electron dynamics in all the listed mechanisms, and end with two points of discussion; first, we can choose a dominant mechanism by appropriately selecting not only the impact energy, but also the impact parameter; and second, calculations from our model predict that non-first order mechanisms (all but \textit{TS1} and \textit{TS1*}) dominate over the first order mechanisms, \textit{TS1} and \textit{TS1*}, which is observed in experiments. Among the non-first order mechanisms, \textit{TS2} and \textit{TS2*}, taken together are most dominant followed by \textit{D1} and \textit{D1*} taken together, followed by \textit{D0} and \textit{Inner Shell Capture}. Both points of discussion may help motivate future experiments in the field of electron impact ionization. The \textit{Delay} and the \textit{Ionized Inner Shell} mechanisms are consequences of electron-electron correlations with the inner shell electron. Therefore, electron-electron correlations with the inner shell electron give rise to the \textit{Delay} and \textit{Ionized Inner Shell} mechanisms, which in turn contribute to building up the knee in the probability of double ionization of Mg by electron impact.

\section*{Acknowledgments}

We would like to thank Simon Berman for constructive feedback during the initial stage of the research. S.M. would like to thank Somya Mittal and Fereshteh Shahmiri for their curiosity and for asking questions about the research, which were helpful in the writing process. The project leading to this research has received funding from the European Union’s Horizon 2020 research and innovation program under the Marie Sk\l{}odowska-Curie Grant agreement No. 734557.

\end{document}